%% file: main.tex
\documentclass[preprint,12pt,3p]{elsarticle}

\usepackage{graphicx}
\graphicspath{ {figures/}{figures/pdf/} }
\usepackage{amssymb}
\usepackage{amsthm}

\usepackage{amsmath}

\usepackage{aas_macros}
\usepackage{hyperref}
\hypersetup{
  colorlinks,
  citecolor=Blue,
  linkcolor=Red,
  urlcolor=Blue}

\usepackage[nameinlink,noabbrev]{cleveref}

\usepackage{epsfig}
\usepackage{pdflscape}					

\usepackage{longtable}
\usepackage{float}
\usepackage{fancyhdr}

\fancypagestyle{mylandscape}{
	\fancyhf{} 
	\fancyfoot{
    \makebox[\textwidth][r]{
      \rlap{\hspace{.75cm}
        \smash{
          \raisebox{4.87in}{
            \rotatebox{90}{\thepage}}}}}}
}

\usepackage{cancel}

\floatstyle{plaintop}
\restylefloat{table}

\usepackage{caption}
\usepackage{subcaption}
\usepackage{afterpage}

\usepackage{amssymb}
\usepackage{amsthm}
\usepackage{amsmath}
\usepackage{xfrac}

\usepackage{color, colortbl}
\definecolor{Gray}{gray}{0.9}

\usepackage{fancyhdr} 
\usepackage[detect-all]{siunitx}

\newcommand{\R}{{R}}
\newcommand{\C}{{C}}

\newcommand{\unitv}[1]{\hat{\vect{#1}}}
\newcommand{\e}{{\unitv{e}}}

\newcommand{\h}{^\mathrm{h}}
\newcommand{\m}{^\mathrm{m}}
\newcommand{\s}{^\mathrm{s}}

\newcommand{\au}{\,\mathrm{au}}
\newcommand{\km}{\,\mathrm{km}}
\newcommand{\meter}{\,\mathrm{m}}
\newcommand{\cm}{\,\mathrm{cm}}
\newcommand{\mm}{\,\mathrm{mm}}

\newcommand{\Myr}{\,\mathrm{Myr}}
\newcommand{\kyr}{\,\mathrm{kyr}}

\newcommand{\hours}{\,\mathrm{hrs}}
\newcommand{\hour}{\,\mathrm{hr}}

\newcommand{\second}{\,\mathrm{s}}

\newcommand{\K}{\,\mathrm{K}}
\newcommand{\J}{\,\mathrm{J}}
\newcommand{\N}{\,\mathrm{N}}

\newcommand{\g}{\,\mathrm{g}}

\newcommand{\W}{\,\mathrm{W}}

\usepackage{lineno}

\biboptions{authoryear,round}

\begin{document}

\begin{frontmatter}


\title{Thermophysical Investigation of Asteroid Surfaces II: Factors Influencing Grain Size}



\author[label1,label2]{Eric M. MacLennan\corref{cor1}}
\ead{eric.maclennan@helsinki.fi}
\ead{mv.helsinki.fi/home/maceric/home.html}

\author[label1,label3]{Joshua P. Emery}

\cortext[cor1]{corresponding author}

\address[label1]{Earth and Planetary Sciences Department, Planetary Geosciences Institute, The University of Tennessee, Knoxville, TN 37996, USA}
\address[label2]{Department of Physics, P.O. Box 64, 00560 University of Helsinki, Finland}
\address[label3]{Department of Physics and Astronomy, Northern Arizona University, NAU Box 6010, Flagstaff, AZ 86011, USA}

\begin{abstract}

Asteroid surfaces are subjected to mechanical weathering processes that result in the development and evolution of regolith. Two proposed mechanisms--impact bombardment and thermal fatigue--have been proposed as viable and dominant weathering processes. Previously, we compiled and estimated thermal inertias of several hundred asteroids (mostly in the main-belt) for which we determined dependencies on temperature, diameter, and rotation period. In this work, we estimate grain sizes of asteroid regoliths from this large thermal inertia dataset using thermal conductivity models. Following our previous work we perform multi-variate linear model fits to the grain size dataset and quantify its dependency on diameter and rotation period. We find that the preferred model indicates that asteroid grain sizes are inversely dependent on object size for $<$10 km asteroids and exhibits no relationship above this size cutoff. Rotation period and grain size show a positive relationship when the rotation period is greater than $\sim$5 hr, and an inverse relationship below this rotation period. These results indicate that both impact weathering and thermal fatigue are relevant regolith evolution mechanisms. We run post-hoc t-tests between spectral groups to infer the influence of composition on regolith grain sizes. We find that M-type (including suspected metal-rich objects) and E-type asteroids have larger grain sizes relative to our population sample, and that P-types have distinctly smaller grains than other groups.

\end{abstract}


\end{frontmatter}


\section{Introduction}\label{S:1}

The study and characterization of asteroid regolith---the unconsolidated, heterogeneous, rocky material covering the surface of planetary bodies \citep{Shoemaker_etal69}---is an important part of understanding the processes and evolution of airless bodies of the Solar System. Generally speaking, asteroid surfaces evolve from poorly-sorted, blocky mixtures to well-sorted fine-grained regolith \citep{Horz_etal20}. Thermal inertia, $\Gamma$, is a thermophysical property that can be used to characterize asteroid regolith. Thermal inertias lower than that of bare rock \citep[$\Gamma \sim 1500$--$2500 \J \meter^{-2} \K^{-1} \second^{-1/2}$][]{Jakosky86,Bandfield_etal11} indicate the presence of regolith, with lower values signaling a finer-grained surface. Although some meteorites have measured thermal inertia lower than this \citep[$\Gamma \sim 1000 \J \meter^{-2} \K^{-1} \second^{-1/2}$;][]{Opeil_etal10,Opeil_etal20}, the range of many asteroid thermal inertias \citep[$< 150 \J \meter^{-2} \K^{-1} \second^{-1/2}$][]{MacLennan+Emery21} is significantly lower. The bulk thermal inertia can be expressed as $\Gamma = \sqrt{k_\mathit{eff} \rho_\mathit{grain} c_s (1-\phi})$ in which $k_\mathit{eff}$ is the effective thermal conductivity of the regolith, $\rho_\mathit{grain}$ is the density of a rock with no void spaces, $c_s$ is the specific heat capacity of the material, and the regolith porosity, $\phi$, is explicitly accounted for. Some mechanisms have been suggested as the primary drivers of regolith mechanical weathering on airless bodies: meteoroid impacts and thermal cycling.

Small meteoroid ($<1 \meter$) impacts can cause breakdown of surface material over time. Some of the energy from an impact is partitioned into fragmenting near-surface material of the target body, albeit the exact fraction of energy that goes into this mechanical work is uncertain \citep{Horz&Cintala97}. Some of the energy is partitioned into launching particles off the surface on various trajectories (ejecta), which partly depend on the proximity to the impact site and on the mechanical properties of the impactor and target. Particles that do not reach escape velocity return to the surface as newly-formed regolith. Other factors, such as the target porosity and strength, also play relevant roles in the production of craters and of regolith. For example, \cite{Housen&Holsapple03} found a clear inverse relationship between the ejecta:impactor mass ratio and porosity of the target. Impacts that excavate into fractured bedrock will generate more ejecta blocks. Thus, regolith evolution driven by impacts is dependent on on the size of the body, strength of the material, among other factors.

Internal stresses caused by differential thermal expansion, as a result of cyclic heating and cooling, lead to regolith breakdown and erosion \citep{Eppes_etal15}. The efficiency of this thermal cycling process has been modeled under the thermal environments of airless solar system bodies \citep[e.g.,][]{Molaro&Byrne12,Molaro_etal17}, and its feasibility has been experimentally demonstrated on meteorite samples \citep{Delbo_etal14,Libourel_etal18}. \citet{Molaro_etal19} demonstrated that large cracks observed on rocks seen across Bennu's surface were consistent with thermal fatigue model predictions. In general, increasing the heterogeneity of a rock (by changing the mineral grain boundaries or pores of empty space) can raise the peak stresses by up to a factor of three, compared to a homogeneous rock \citep{Molaro_etal15}. Thermal fracturing acts on spatial scales spanning many orders of magnitude, and modeling work has shown that the efficiency depends on heliocentric distance, rotation period, and material thermomechanical properties \citep{Ravaji_etal19,ElMir_etal19}. The propagation of the thermal wave within a boulder initiates one or more cracks at the micro-scale throughout the boulder interior and throughout the diurnal cycle \citep{Hazeli_etal18}. These micro-cracks most likely originate near mineral grain boundaries, preferentially grow perpendicular to the local surface, and ultimately intersect other fractures reaching sizes of several centimeters \citep{Delbo_etal14,Molaro_etal15}. Growth of large-scale cracks occurs in the direction of the heat flow throughout the boulder, and are thus able to transverse rocks that are several times larger than the initial fracture(s). Although rapidly rotating asteroids exhibit small diurnal temperature differences, \cite{Delbo_etal14} showed that even a short heating cycle of $2.2~\hours$ is sufficient to generate and grow cracks. Lastly, \cite{Molaro&Byrne12} claim that fast-rotators ought to facilitate larger maximum thermal gradients, thus predicting a rotation period dependence of the overall efficiency.


This work (part II) continues our thermophysical investigation of asteroid surface properties by estimating thermally-characteristic grain sizes from thermal inertias \citep[part I;][]{MacLennan+Emery21}. We use these grain sizes to study regolith evolution in the context of two proposed regolith evolution mechanisms: meteoroid impact degradation and thermal fatigue cycling. These processes have been proposed as relevant for the creation and subsequent evolution of regolith on asteroid surfaces. The relevance of each weathering mechanism can be examined by comparing the relative correlation between regolith grain size and asteroid diameter or rotation period. Specifically, we claim that if regolith development and evolution is highly dependent on meteoroid impacts then we expect a statistically-significant correlation between the grain sizes and asteroid diameters. Similarly, if thermal cycling is effective then we should observe that grain sizes are correlated with rotation periods. These two mechanisms are not mutually exclusive, thus we consider the possibility that both are relevant for asteroid surfaces by using a multi-variate linear model in our analysis (\autoref{subsec:MVR}). We use this multi-variate model to test the following hypotheses: 1) larger asteroids exhibit evolved regoliths, quantified by smaller grain sizes, and 2) slowly-rotating asteroids have a poorly-developed regolith, characterized by larger grain sizes. We also investigate how the regolith properties of main-belt asteroids are related to spectral classification using post-hoc tests of the fitted model residuals (\autoref{subsec:compgrainsize}), and orbital properties of near-Earth asteroids (\autoref{subsec:NEAanalysis}). Finally, we note that it is currently unclear in what ways these grain sizes characterize a heterogeneous mixture of e.g., regolith grains and boulders. Thus, although the grain sizes estimated herein may be inaccurate in the absolute sense (\autoref{subsec:datalimit}) we utilize the modeling results to interpret the trends (\autoref{sec:discussion}) within the asteroid population in the context of regolith evolution mechanisms.

\section{Methods}\label{sec:grainmodel}

A handful of models have been developed for estimating grain size from thermal inertia of planetary surfaces by experimentally observing and/or theoretically modeling heat flow in granular media. For example, laboratory experiments have been able to observe empirical effects of particle size on $k_\mathit{eff}$ \citep{Watson64,Presley&Christiansen97,Presley&Craddock06,Sakatani_etal18}. Since, for airless planetary regoliths, $\Gamma$ is primarily, but not exclusively, influenced by changes in $k_\mathit{eff}$, an estimate of grain size can be made from $\Gamma$ by using an appropriate thermal conductivity model for a granular medium \citep[e.g.,][]{Gundlach&Blum13,Sakatani_etal17,WoodsRobinson_etal19,Wood20}. In general, for airless bodies $k_\mathit{eff}$ has a solid-state component that describes heat conduction through grains and across grain contacts, and a radiative component that describes heat radiated across pore spaces \citep[e.g.,][]{Watson64}:
\begin{equation}\label{eq:keff}
	k_\mathit{eff} = k_\mathit{solid} + k_\mathit{rad} = k_1(T) + k_2T^3.
\end{equation}
The $k_1$ and $k_2$ coefficients in \autoref{eq:keff} are dependent on the material properties of the regolith, such as grain size, packing fraction, and amount of contact between the grains \citep{Watson64}. Note that the solid component of thermal conductivity is temperature-dependent and varies with composition \citep[][and references therein]{Wood20}. These coefficients are approximated in the works of \cite{Gundlach&Blum13} and \cite{Sakatani_etal17}, from which measurements of $\Gamma$ (combined with the compositional information about the asteroid) can be used to estimate a characteristic grain size for an asteroid. We thus employ both these models, with an anisothermality correction from \citet{Ryan_etal20} applied to the radiative conductivity term, to estimate characteristic regolith grain sizes for each asteroid.

The \cite{Gundlach&Blum13} formulation of $k_1$ and $k_2$ are calibrated using laboratory heat-flow measurements of lunar regolith \citep{Chan&Tien73,Gundlach&Blum12}. On the other hand, \citet{Sakatani_etal17} used mixtures of powdered glass beads of varying sizes and porosities. The surface temperature is required input for these models, from which we use the color temperature reported in Part I \citep{MacLennan+Emery21}. These models also need compositional information about the material, which we infer from the asteroid's spectral taxonomic type when available. For objects for which there is no spectral information available, we infer the spectral type from the geometric albedo. Many asteroids belong to a dynamical family, in which case we can infer the spectral type and perform a cross-check with the geometric albedo or reflectance spectrum, if available. We assign a meteorite analog to each spectral type as outlined in \autoref{subsec:2.22} and account for uncertainties in the thermal inertia and various material properties by using a Monte Carlo method (\autoref{subsec:MC}) when implementing the model.

\subsection{Thermal Conductivity Modeling}\label{subsec:kmodel}

The solid thermal conductivity component, $k_1$, is modeled by \citet{Gundlach&Blum13} via computing the efficiency of heat transfer through the surface contacts within a network of grains:
\begin{equation}\label{eq2.8}
	k^{G\&B}_1 = k_\mathit{grain} \frac{r_c}{r_g} (f_1\ \exp([f_2 \psi]))\ \Xi.
\end{equation}
$k_\mathit{grain}$ is the thermal conductivity of a grains with zero porosity, $\psi = 1 - \phi$ is the volume filling factor (packing fraction). The empirically-derived constants $f_1 = (5.18\ \pm\ 3.45) \times 10^2$ and $f_2 = 5.26\ \pm\ 0.94$ encapsulate information about the path of contact chains in the direction of heat flow \citep{Gundlach&Blum12}. Thermal conductivity measurements of Apollo 11 \& 12 samples are used to estimate $\Xi = 0.41 \pm 0.02$, which incorporates and accounts for the irregular shapes of the particles and heterogeneity of regolith on the whole. The factors contained within brackets in \autoref{eq2.8} approximate the adhesive forces between regolith grains, which dominate over gravity on small bodies, to estimate the contact area between them. The contact radius between grains, $r_c$, is calculated by JKR (Johnson-Kendall-Roberts) theory \citep{Johnson_etal71}, assuming that adhesive forces are dominant:
\begin{equation}\label{eq:contactr}
    r_c = \Bigg[ \frac{9\pi}{4} \frac{(1-\nu^2)}{E} \gamma(T) r^2_g \Bigg]^{-1/3}.
\end{equation}
where $\nu$ is Poisson's ratio, $E$ is Young's Modulus, and $\gamma(T) = T\cdot 6.67 \times 10^{-5} J m^{-2}$ is the specific surface energy of each grain---a measure for the adhesive bonding strength between grains.

In \citet{Sakatani_etal17}, the solid conductivity component is modeled in a similar way to the formula above:
\begin{equation}
    k^{Sak}_1 = k_\mathit{grain} \frac{r_c}{r_g} \frac{4\psi C \xi}{\pi^2}
\end{equation}
where $r_c$ is the contact radius, also represented by \autoref{eq:contactr}. The coordination number,
\begin{equation}
    C = \frac{2.8112 (1-\phi)^{-1/3}}{f^2(1+f^2)},
\end{equation}
is the average number of particles that are in contact with each other and is a function of the regolith porosity. Here, $f = 0.07318 + 2.193\phi - 3.357 \phi^2 + 3.914\phi^3$. The factor $\xi$ is dependent on the shape and smoothness of the particles and is equal to unity for perfectly smooth spheres. We use $\xi = 0.4$, which approximates rough, non-spherical particles and is well within the range of experimentally-derived values from \cite{Sakatani_etal17}.

The radiative thermal conductivity ($k_\mathit{rad}$) coefficient is calculated by \cite{Gundlach&Blum13} to be:
\begin{equation}\label{eq:GB2}
	k^\mathrm{G\&B}_2 = 8 \epsilon \sigma_0 e_1 \frac{1-\psi}{\psi} r_g,
\end{equation}
with the bolometric emissivity, $\epsilon$, Stefan-Boltzmann constant, $\sigma_0 \approx 5.67 10^{-8} \W \meter^{-2} \K^{-4}$, and the empirically-derived coefficient $e_1 = 1.34 \pm 0.01$ \citep{Dullien91,Gundlach&Blum12}. The mean free path of a photon between regolith grains is directly proportional to their size, hence the $r_g$ factor in \autoref{eq:GB2}. In \cite{Sakatani_etal17} the radiative heat transfer coefficient is calculated via:
\begin{equation}\label{eq:Sak2}
    k^{Sak}_2 = 8\sigma_0 \zeta \frac{\epsilon}{2-\epsilon} \Big[ \frac{1-\psi}{\psi}\Big]^{1/3} r_g
\end{equation}
where $\zeta$ is an enhancement factor for which we use $\zeta = 1.7$---an average of the experimentally-derived values in  \citet{Sakatani_etal17}. However, we note that $k_\mathit{rad}$ values presented in \citet{Sakatani_etal17} show a clear inverse dependency on $r_g$, which may be related to the effect of short-range (immediately-adjacent) versus long-range (non-adjacent) radiation exchange that is dependent on the particle size \citep{vanAntwerpen_etal12}.

Thermal gradients that exist within individual regolith particles cause a non-linear relationship between $r_g$ and $k_\mathit{rad}$ \citep{Ryan_etal20}. Such thermal gradients will arise in large particles, when the solid component of thermal conductivity is low, or for particles that have significant intra-granular porosity. In these cases, the radiative component of thermal conductivity will be less than the $T^3$ theoretical relationship. Building off the work of \citet{vanAntwerpen_etal12}, \citet{Ryan_etal20} use the dimensionless parameter $\Lambda_s$ in an updated version of the non-isothermal correction factor, $f_k$:
\begin{equation}
    \Lambda_s = \frac{k_{grain}}{8r_g\sigma_0T^3}
\end{equation}
and
\begin{equation}\label{eq:fkorr}
    f_k = a_1 \tan^{-1}(a_2\Lambda_s^{-a_3}) + a_4.
\end{equation}
When $\Lambda_s > $25 then $f_k$ is set to unity, otherwise $f_k$ can be calculated via \autoref{eq:fkorr} where a$_1$=-0.568, a$_2$=0.912, a$_3$=0.765, and a$_4$=1.035. To account for this non-isothermality, we multiply the coefficients calculated in \autoref{eq:GB2} and \autoref{eq:Sak2} by $f_k$.

\subsection{Input Parameters}\label{subsec:2.22}

Some of the required input parameters to the equations listed above are dependent on the composition of the material. Since meteorites are samples of asteroids, we attempt to establish a meteorite analog that is most appropriate for each asteroid in this study. The most direct approach can be made when high quality spectra have been acquired for an object. Various taxonomic systems have been defined based photometric colors and albedos \citep{Tholen84}, or absorption features with spectral slopes of reflectance spectra \citep{Bus&Binzel02b,DeMeo_etal09}. The taxa defined in these works relate to different compositions, although connections can be ambiguous in many cases; particularly, for featureless spectra. Visible spectra and color information can be used to broadly distinguish between the S-complex (any spectral taxon beginning with an ``S'' and Q-types that are the unweathered endmember), K-type, V-type, B-type, C-complex (any taxon beginning with a ``C''), and Bus-DeMeo X-complex. For the X-complex we adopt the \citet{Tholen84} E, M, and P-type system, which are distinguished by their geometric albedos. We use V-band geometric albedo cutoff of $p_V \gtrsim$ 0.42 for E-type objects, 0.12 $< p_V <$ 0.42 for M-types, and $p_V \lesssim$ 0.12 for P-types. 

Many asteroids in our sample have no color or spectral information available, thus we are left to infer their compositional information by other means. For this exercise we use the geometric albedo reported and compiled by \citet{MacLennan+Emery21}, and V-band slope parameters ($G_V$) from \citet{Oszkiewicz_etal11}, which have both been shown to correlate with spectral class \citep[e.g.][]{Oszkiewicz_etal12,Veres_etal15}. We found that using cutoff criteria only for $p_V$ was sufficient, but still utilize $G_V$ values as a consistency check that each object is within the range of expected values for its spectral class. All objects' $p_V$ and $G_V$ values are included in \autoref{fig:albG}. It is interesting to note the positive correlation between $p_V$ and $G_V$ for the entire dataset (across taxonomic groups), yet, within each taxonomic group this correlation is absent or exhibits an inverse relationship. Such a scenario is an example of Simpson's statistical paradox\footnote{Simpson's paradox describes a situation in which a dataset exhibits correlations across the entire sample that are statistically distinguishable from sample correlations within groups.} \citep{Simpson51,Yule1903}. We briefly note this paradox for possible future investigations of $p_V$, $G_V$, and regolith grain-scale size across different spectral/compositional groups.

To distinguish between high-albedo S-complex and low-albedo C-complex we use $p_V = 0.12$, whereas objects with $p_V > 0.45$ are assumed to be E-types. The $p_V = 0.12$ cutoff value is consistent with \cite{DeMeo&Carry13}, who present average albedos for different asteroid taxa using a larger sample size than ours. This cutoff criterion is also consistent with laboratory-derived geometric albedos of ordinary chondrites and CO, CM, and CI carbonaceous chondrites \citep[fig. 7 of][]{Beck_etal21}. However, the laboratory-derived values for CK and CV chondrites have significant overlap with unequilibrated (type 3) ordinary chondrites for $0.12 < p_V < 0.17$. Thus, with this criteria, meteorite analog associations for 8 objects in our sample with albedos in this range are considered to be ambiguous.

We assign ordinary chondrite and carbonaceous chondrite material properties to S-complex (including one O-type) and C-complex (excluding Ch-type) asteroids, respectively. We associate the low-density CM chondrites with Ch and B-type, and CI chondrites with P-type asteroids. Our object set contains a few D-types (color or spectral classification) for which we assume a P-type composition that most likely reflects the primitive compositions of these asteroids. Asteroids classified as K-type (and Xk-type) have been shown to be related to CO, CV, and CK meteorites \citep{Burbine_etal01b,Clark_etal09}, and can mostly be found among the Eos family \citep{MotheDiniz_etal08} in our sample. The V-type association to the HEDs, a basaltic a chondrite meteorite class, has been well established. E-type (Tholen) and Xe-type (Bus-DeMeo) asteroids have a well-established connection to aubrites (enstatite achondrites) because of their high albedos.

Enstatite chondrites, which have a distinct thermal conductivity compared to other meteorite groups (\autoref{fig:kheatcap}) have been suggested as an analog to the M-types \citet{Clark_etal04}. On the other hand, M-types that have a high radar albedo and thermal inertia are suggestive of a relatively high FeNi metal content \citep[e.g.][]{Magri_etal99}. Spectroscopic modeling of the largest M-type, (16) Psyche, indicate a regolith comprised of a silicate-metal mixture with exogenous carbonaceous material \citep{Landsman_etal18,Cantillo_etal21}---the former being consistent with results smaller M-types \citep{Sanchez_etal21}. The Psyche mission \citep{ElkinsTanton16} should reconcile the various thermal inertia estimates\footnote{We clarify that we have chosen to use Psyche's thermal inertia that was obtained by \citep{Matter_etal13}, which is based on disk-resolved, mid-infrared observations--a wavelength range that is comparable to most other thermal inertias used for this work.} that range from $<$100 J m$^{-2}$K$^{-1}$s$^{-1/2}$ \citep{Landsman_etal18} up to $210 \pm 60$ J m$^{-2}$K$^{-1}$s$^{-1/2}$ \citep{deKleer_etal21} with the spectroscopic evidence, and provide crucial information (i.e., porosity, emissivity, and metal content) that can be used to model the surfaces of other suspected metal-rich M-types. Although we assume a purely metal surface for all suspected metal-rich asteroids, which is most likely an inaccurate assumption, we show in \autoref{subsec:datalimit} that this assumption does not drastically affect the grain size estimate.

The thermal conductivity ($k_\mathit{grain}$), grain density ($\rho_\mathit{grain}$), heat capacity ($c_s$), Young's modulus ($E$), and Poisson's ratio ($\mu$) for all aforementioned groups are shown in \autoref{tab:table2.2}. In the following subsections, we describe how these properties were chosen for each spectral group and analyze data from many sources, when appropriate, to account for temperature and porosity effects. We note that, in our albedo-based classification, M-types and low-albedo V-types could potentially be misclassified as S-types and, similarly, some P-types may be misclassified as C-types. However, we see no obvious reason to re-assign any of these S-types to the M-type group, as their thermal inertias are inconsistent with a high thermal conductivity which would otherwise suggest a metal-rich surface. On the other hand, some P-types might be misclassified as C-types. We change the classification of (4003) Schumann from C-type to P-type, based on its location in the outer part of the Main Belt ($\approx3.4 \au$) where P-types are more abundant over C-types \citep{DeMeo&Carry13,DeMeo&Carry14}. The difference in assumed material properties between C and P-types that we use is not large and would not alter the reported grain size by more than a few percent.

\subsubsection{Thermal Conductivity} Laboratory measurements of meteorite thermal conductivities reveal dependence on the material porosity and temperature. In their review paper of thermal conductivites, \cite{Flynn_etal18} show that laboratory measurements of thermal conductivity do not change significantly in the temperature range from 100 and $300\K$, but approach zero at $0\K$. Additionally, void spaces can significantly impede the net solid-state heat flow within a meteorite sample \citep{Soini_etal20}. To account for porosity effects, we estimate the thermal conductivity at zero porosity using data collected at the same temperature.

In order to estimate thermal conductivity at $\phi = 0$, lines are fit to meteorite samples with porosities less than $12\%$ from the literature (\autoref{tab:porosityk}), as shown in \autoref{fig:kheatcap}. Since the carbonaceous chondrites with measured $k_\mathit{eff}$ had porosities exceeding this cutoff, we did not use any for our analysis. We found that enstatite chondrites have a larger thermal conductivity than the ordinary chondrite samples (and one Shergottite), likely due to mineralogical differences. The $y$-intercept of the best-fit lines give the thermal conductivity at zero porosity of 4.05 and 6.23 W $\meter^{-1} \K^{-1}$ for ordinary chondrites and enstatite chondrites, respectively. Because the thermal conductivity for terrestrial enstatite is $90\%$ that of enstatite chondrites at $275 \K$ we scale the function downward by 10\% to approximate the aubrite thermal conductivity, which are mostly comprised of enstatite. We used the same line slope to extrapolate a single measurement of thermal conductivity \citep{Opeil_etal10} of the FeNi meteorite Campo del Cielo to estimate the thermal conductivity at $\phi = 0$ are used in the thermal conductivity model to compute grain size.

Data presented by \citet{Soini_etal20} shows that the thermal conductivities of ordinary and carbonaceous chondrites are in agreement when the sample porosities are taken into account. Because we account for sample porosities and are using grain densities, we use the same zero-porosity thermal conductivity value for carbonaceous chondrites and ordinary chondrites, with the exception of CM chondrites. Newer thermal conductivity measurements of CM chondrites from \cite{Opeil_etal20} are also consistent with other carbonaceous chondrites. However, the spacecraft observations of B-type Bennu and Cg-type Ryugu revealed surfaces with weak, low-conductivity boulders that have thermal inertias which are $\sim$2--3 times smaller than stronger, high-conductivity boulders \citep{Grott_etal19,Rozitis_etal20}. This stronger material may be identical to CM chondrites found on Earth, yet, the weak material is likely less abundant among meteorite collections \citep{Popova_etal11}. In order to account for the presence of this weak material that is likely present on Ch and B-type asteroids, we adopt a lower thermal conductivity for both. Because this weaker material is most likely primitive in origin, we assign the low thermal conductivity to CI chondrites as well.

Our computed thermal conductivity values agree with the \citet{Soini_etal20} model fits to essentially the same dataset as ours. We accounted for temperature-dependent thermal conductivity among these samples by bootstrapping our $k_\mathit{grain}$ calculations with the data compiled in \citet{Flynn_etal18}. Ordinary chondrites, HEDs, and carbonaceous chondrites do not demonstrate a significant dependence on temperature \citep{Opeil_etal12,Flynn_etal18} in the range relevant to this study, so we use constant values. For FeNi and enstatite chondrite curves, we scale the curves presented in \citet[Fig. 15;][]{Flynn_etal18} to our values of $k_\mathit{grain}$ at $200\K$. FeNi meteorites show a roughly linear dependence, and enstatite chondrites (and by deduction, the enstatite-dominated aubrites) show an inverse dependence on temperature. These temperature-dependent $k_\mathit{grain}$ functions, which are valid for 200--$350\K$, are listed in \autoref{tab:table2.2} and are assigned a 10\% uncertainty in the thermal conductivity model.

\subsubsection{Grain Density and Heat Capacity} Meteorite specimens, while not regolith, still contain small pores and cracks that lower the bulk density of the sample \citep[e.g.,][]{Cadenhead&Stetter75}. These micro-porosities, $\phi$, range from 4\% to 10\%, but can be as high as 20\% for weathered fines and as low as 2\% for iron meteorites. We wish to use the grain density ($\rho_\mathit{grain}$), the density of a material with zero porosity, as input to the thermal conductivity models. We primarily use the findings \cite{Macke10}, who measured the porosities of individual meteorites with various compositions and computed $\rho_\mathit{grain}$. \autoref{tab:table2.2} lists grain densities for many meteorite analogs related to their respective spectral types. For the S-complex and V-types we use a uniform distribution for $\rho_\mathit{grain}$ that spans the values reported in \cite{Macke10}, instead of a normal distribution.

The ability of material to store thermal energy per unit mass is quantified by its specific heat capacity, $c_s$. Laboratory heat capacity measurements of several meteorites reveal a clear temperature-dependence  \citep{Beech_etal09,Opeil_etal12,Szurgot_etal12,Consolmagno_etal13,Wach_etal13,Macke_etal19,Opeil_etal20}. We account for this dependence by performing an independent meta-analysis---similar to that on the thermal conductivity dataset above---in which we use literature data on meteorite heat capacities measured at various ambient temperatures. While \cite{Flynn_etal18} perform a similar analysis, they present fits for a temperature range of 75--$200 \K$, which does not cover the full range of asteroid surface temperatures relevant to this work. Our empirical fits to the data are in agreement to the semi-empirical functions of \cite{Macke_etal19}, who used a larger dataset. Our results are very similar to theirs over the same temperature range.

\citet{Opeil_etal20} present temperature-dependent heat capacities for five CM chondrites, which we compare our results to below. We compute $2nd$ order polynomial fits to the data shown in \autoref{tab:heatcapT} over a temperature range of 175--$300\K$ and with a $y$-intercept fixed at zero (because at low temperatures the heat capacity approaches zero). Among our dataset, we find that two distinct trends emerge, forming two groupings: one with all the iron-nickel meteorites (metallic) and one comprised of all chondritic and achondritic meteorites (non-metallic). The best-fit heat capacity equations (and coefficient uncertainties) for non-metallic and metallic meteorites at different temperatures are given by $c_s$ = -0.0033 ($\pm$ 0.0004)$\times T^2$ + 3.39 ($\pm$ 0.10)$\times T$ and $c_s$ = -0.0044 ($\pm$ 0.0004)$\times T^2$ + 2.84 ($\pm$ 0.11)$\times T$, respectively. Results are shown in \autoref{fig:kheatcap} and the fits from \cite{Macke_etal19} and \citet{Opeil_etal20} are included for comparison. Because the \citet{Opeil_etal20} CM chondrite dataset clearly exhibit higher heat capacities than our fits to other non-metallic meteorites, we use the function $c_s$ = -0.0036$\times T^2$ + 3.84$\times T$ for Ch and B-type asteroids, which is calculated as 115\% of the fit to the non-metallic meteorites. As shown in \autoref{fig:kheatcap}, this scaled function falls in the range of formulas presented by \citet{Opeil_etal20}.

\subsubsection{Mechanical Properties and Emissivity} Young's modulus ($E$) and Poisson's ratio ($\nu$) are used in the thermal conductivity model to calculate the contact area between grains within a regolith. In particular, these two quantities are used to approximate the deformation both along the axis of an applied force and in the orthogonal dimensions. For M-type asteroids, we use the values of these properties measured for Fe-Ni alloys, which are similar to iron meteorites, by \cite{Ledbetter&Reed73}. \cite{Ibrahim12} reports Young's and Bulk Modulus ($G$) for many ordinary and carbonaceous chondrite meteorites. We calculate Poisson's Ratio using the relationship between the three variables: $\nu = \frac{E}{2G} - 1$. This equation assumes isotropic material properties and is also used to propagate the reported $E$ and $G$ uncertainties presented in \cite{Ibrahim12}. The values for E-type asteroids are assumed to be the same as used for S-types, as indicated by italics in \autoref{tab:table2.2}.

Emissivity values of meteorites and terrestrial rocks cluster around 0.9--the value that we adopt in this work for all spectral groups with the exception of suspected metal-rich asteroids. Laboratory and modeling efforts by \citet{Sih&Barlow04} place the emissivity of powdered and solid iron roughly in the range of 0.5 to 0.8 for temperatures of 200 to $300\K$. \citet{Gundlach&Blum13} assume an emissivity for metal-rich asteroids of 0.66, which we adopt here.

\subsection{Monte Carlo Implementation}\label{subsec:MC}

We compute a thermal conductivity, $k\mathit{^{obs}_\mathit{eff}}$, that is empirically-derived from $\Gamma$ and assumed values of $c_s$, $\rho_\mathit{grain}$, and a range of porosities. This value is equated to the theoretically-modeled thermal conductivity calculated using the above procedure, in order to obtain an estimated regolith grain size. Most input variables are taken from a distribution that is generated based on the uncertainty of that parameter given in \autoref{tab:table2.2} and mentioned above. Most parameters have associated 1$\sigma$ uncertainties from which we generate Gaussian probability distributions. The porosity is one exception, for which we use a uniform random distribution from 0.276 to 0.876; the lower value is the porosity of close-hexagonal packing scheme for identical spheres and the upper value represents the hypothesized porosity of cometary regolith \citep[i.e.,][and references therein]{Sunshine_etal16}. For each iteration, a single grain size is produced, ultimately constructing a distributed set of 10,000 grain sizes. For some trials, the combination of input parameters produced model thermal conductivity values that are incompatible with the observed value. These cases are therefore not included in the final distribution and are instead replaced by additional, successful trials. We note that failed trails are indicative of input parameters that fall at the tail ends of the adopted distributions and do not necessarily imply incorrect assumptions with the models.

Output grain sizes are transformed into the logarithmic scale developed by \cite{Krumbien&Aberdeen37} which allows for the comparison of sediment sizes across many orders of magnitude. These Krumbien {\it phi-scale} grain sizes, $d_{g\phi}$, are referenced from $1\mm$. The output grain diameters, $d_g = 2 r_g$, are transformed into this scale with:
\begin{equation}
	d_{g\phi} = -2\ \textrm{log}_2(d_g/\textrm{1 mm}).
\end{equation}
Note that smaller (and negative) values on this scale represent larger grain sizes. Output grain size distribution for an object does not necessarily, or even typically, represent a Gaussian distribution. Thus, we report the median of this distribution as the best-fit grain size, and report upper and lower uncertainties by computing the grain sizes that encompass one standard deviation from the median value. The diurnal e-folding thermal skin depth\footnote{Both \cite{MacLennan+Emery19} and \cite{MacLennan+Emery21} incorrectly quote the formula for $l_s$ by including a factor of $1/\sqrt{2}$, although no values were reported in either work.},
\begin{equation}\label{eq:skindepth}
l_s = \sqrt{\frac{k_\mathit{eff}P_{rot}}{\pi c_s \rho_{grain}(1-\phi)}} = \sqrt{\frac{P_{rot}}{\pi}} \frac{\Gamma}{c_s\rho_\mathit{grain}(1-\phi)}
\end{equation}
is also computed and the median and standard deviations from the output distribution are reported.

\subsection{Caveats and Model Limitations}\label{subsec:datalimit}

For some asteroids in our sample, the material properties inferred from spectroscopy, color, or albedo data may not be correct. For example, incorrect assignment of an asteroid as a metallic M-type when it is actually comprised of silicate S-type material is a possible case that could cause non-negligible change in the grain size estimate. To understand the effects of incorrect classification, we modeled the thermal conductivity model for (22) Kalliope (an M-type) with both metal-rich and S-type properties. Kalliope's low radar albedo may suggest a silicate-rich instead of a metal-dominated surface \citep{Lupishko&Belskaya89,Ockert-Bell_etal10,Hardersen_etal11}. Assuming that Kalliope has S-type properties instead of being metal-rich increased the grain size by $\sim$8\%, which is {\it far} below the typical $d_{g\phi}$ uncertainty reported in \autoref{grainsizeresults}. From this case, we can reasonably assume that uncertainty in the assumed material properties do not significantly contribute any significant bias in grain sizes. 

Both of the thermal conductivity models assume homogeneous, monodispersed grain sizes throughout the surface and constant thermophysical properties with depth. This is also true of reported thermal inertia values. Surface processes that sort grain sizes both vertically and spatially are likely present on asteroids \citep[][and references therein]{Richardson_etal20}. Thus, this assumption of regolith homogeneity is certainly not the case for \emph{any} asteroid, and many spacecraft missions have revealed surfaces that are heterogeneous (\autoref{fig:ErosIto}). Specifically, the Hayabusa mission showed that the surface of Itokawa was consistent with the remotely-determined thermal inertia (600--$800 \J \meter^{-2} \K ^{-1}\second^{-1/2}$), yet still contained regions dominated by fine-grained regolith---counter to interpretation of a high thermal inertia \citep{Muller_etal14,Yano_etal06}. Overall, most asteroid regoliths are expected to be heterogeneous, and the assumption of a single grain size and homogeneous regolith is a practical simplification. However, estimates on the spatial heterogeneity of an asteroid surface are limited due to the fact that we are using disk-integrated (spatially-unresolved) observations.

The recent spacecraft visits to two primitive NEAs, Ryugu and Bennu, unexpectedly revealed surfaces that were different from the initial interpretation of their respective thermal remote observations (see \autoref{fig:ErosIto}). The surfaces of these objects exhibited surfaces comprised of a significant fraction of large porous boulders that were significantly larger than the grain size estimated from thermal inertia \citep{Grott_etal19,Rozitis_etal20}. Because these boulders have a low thermal conductivity that is similar to a coarse-grained regolith, they exhibit thermal properties indistinguishable from a coarse-grained regolith. \cite{Ryan_etal20} claimed that the underestimated grain size of Bennu from its thermal inertia can be explained by the inherently low thermal conductivity of Bennu's surface, which resembles CM chondrites and causes non-isothermal effects within the regolith. Grain size estimates are larger when accounting for this non-isothermality, thus can partly explain the large boulders on Bennu and Ryugu.


Because asteroid regoliths are assuredly a mixture of particle sizes the reported grain size estimates herein should be thought of as a \emph{thermally-characteristic} grain size. Furthermore, it is unknown how this grain size relates to the size distribution of regoloith particles. \cite{Presley&Craddock06} modeled the thermal conductivity of known granular mixtures and deduced that the modeled grain size is most representative of the larger regolith grains---specifically the $85th$ to $95th$ percentiles---rather than the mean or modal grain size. On the other hand, \cite{Ryan_etal20} showed that the particle size from thermal inertia corresponds to within $15\%$ of the mean volumetric particle size for various size-frequency distributions. Additionally, because emitted thermal flux is strongly temperature dependent ($\propto T^4$) it is possible that these thermally-characteristic grain sizes are more representative of warmer regolith patches that consist of smaller grains. Another complication could exist due to the fact that warmer areas emit flux at shorter wavelengths as opposed to cooler areas that emit flux at longer wavelengths. Therefore, its possible that the thermal inertia and thermally-characteristic grain sizes are correlated with the wavelength(s) used in the observation(s). Investigation into the relationship between the grain size and regolith particle distribution is beyond this work and we suggest it as a topic for future studies.

As remarked by \citet{Ryan_etal20}, the apparent thermal inertia will be that of solid rock when the effective particle size is approximately larger than the skin depth, which could be the case for fast rotators with low $k_\mathit{grain}$. If the size range of particles exceeds the skin depth, then the surface is best modeled as a lateral, checkerboard-like mixture of regolith and bare-rock (boulders) \citep[e.g.,][]{Bandfield_etal11,Rozitis_etal20}. Despite the caveats, limitations, and uncertainties of the thermal conductivity model, we claim that the grain sizes derived here are appropriate to identify potential trends among asteroid surfaces.

\section{Results and Analysis}\label{grainsizeresults}

We report modeled estimates of grain size and thermal skin depth in \autoref{tab:table2.3}, along with the model input parameters (with associated uncertainties) for each object---$\Gamma$, $T_\mathit{surf}$ and spectral type. In some cases, the estimated grain sizes exceed the calculated thermal skin depth (see \autoref{subsec:skindepth}). In order to caution readers we mark grain sizes that exceed the lower 1$\sigma$ skin depth lower limit with a ``$^\star$'' in \autoref{tab:table2.3}. We find that the \cite{Sakatani_etal17} thermal conductivity model is incompatible with some of the thermal inertias in which the model over estimates the thermal conductivity. In such cases, only the grain size from the \citet{Gundlach&Blum13} model are reported. Similarly, we find that several C-type asteroids have very low reported thermal inertias that were incompatible with the assumed thermal conductivity. In these cases, we re-ran the model with B-type input parameters, which assume a lower value for the thermal conductivity. We mark these objects as ``B$^\star$'', and include them in the assumed B-type group for the post-hoc analyses. Lastly, we have found that the albedos of a few objects in our set that are inconsistent with the spectral type of the reported family. Similar to \cite{Masiero_etal13}, we thus reject the family membership of these objects and indicate this by crossing through the family name in the rightmost column of \autoref{tab:table2.3}.

We now aim to investigate possible explanatory factors and quantify their influence on the regolith grain size for our sub-sample of the asteroid population. We test the hypothesis that grain sizes are negatively correlated with asteroid diameter (indicating impact driven processes) and whether grain size is positively correlated with rotation period (which is caused by thermal fracturing processes). Various multiple linear (hereafter, multi-linear) regression models are fit to the MBAs to identify and characterize the dependencies of these factors (\autoref{subsec:MVR}). Although the NEAs in this work offer insight into the regolith of very small asteroids, the chaotic nature of changes in their orbital parameters makes it difficult to hold other factors constant---such as thermal environment and impact flux---potentially complicating the multi-linear analysis. We perform a separate analyses on NEAs in \autoref{subsec:NEAanalysis}. In our trend analyses, we use the grain sizes produced by the \citet{Gundlach&Blum13} thermal conductivity model because the \citet{Sakatani_etal17} model was unable to provide an estimate for some objects. In general the \citet{Sakatani_etal17} grain size estimates are consistently larger than those from the \citet{Gundlach&Blum13} model, but we claim that the trends we are investigating should still be found among the \citet{Sakatani_etal17} grain size dataset.

\subsection{Thermal Skin Depth}\label{subsec:skindepth}
First we review the results for the thermal skin depth, $l_s$ of each asteroid in order to place our grain size estimates in better context. As we have mentioned, asteroid surfaces are comprised of a unknown mixture of regolith and boulders of various sizes, and it is unknown how the thermal inertia (and characteristic grain size) are representative of this mixture. It is often thought that $l_s$ ought to be considered, in effect, as an upper limit to the size of regolith grains that can be distinguished from bedrock and that particles that are larger than this have effective thermal inertias equal to bedrock (or a boulder that is much larger than $l_s$). Taking this statement to be truth, we can expect that asteroids with larger $l_s$ should have lower thermal inertias, on average, than those with smaller $l_s$. However, this expectation is confounded by the fact that rotation period also influences $l_s$ (\autoref{eq:skindepth}). To aid our understanding of the relationship between $l_s$, $P_\mathit{rot}$, and $\Gamma$ we place the objects in our dataset into 5 rotation period bins ($< 5\hours$, $5-10\hours$, $10-20\hours$, $20-40\hours$, and $>40\hours$). These rotation period bins are assigned different colors in all three panels of \autoref{fig:skindepth}. The top panel in this figure shows the distributions of $l_s$ for each rotation period bin. The middle and bottom panels show the thermal inertias and calculated grain sizes as a function of skin depth.

When comparing across the rotation period groups in the top panel of \autoref{fig:skindepth} we can see that the two lowest $P_{rot}$ groups have indistinguishable $l_s$ distributions. For the other groups, there is a clear increase in the average $l_s$ among larger rotation periods. Considering this trend, we claim asteroids with $P_\mathit{rot} < 5\hours$ have larger than expected skin depths. The cause for this is lower thermal inertias among the fast rotators, which is shown in \autoref{fig:modelfit} and quantified in our multi-variate model fits in \autoref{subsec:MVR}. The question now becomes: is the higher thermal inertia among fast rotators a result of shrinking $l_s$, or caused by an increase in the regolith grain size?

Comparing the thermal inertias and skin depths (middle panel of \autoref{fig:skindepth}) within the rotation period bins, some interesting findings emerge. Among asteroids with $P_{rot}>40\hours$, the relationship between thermal inertia and skin depth is roughly what is expected by the proportionality relationship implied by \autoref{eq:skindepth} which is shown as a dashed line with an arbitrarily-set intercept. Most of the objects with $l_s > 1\meter$ have very long rotation periods and this subset of objects, on average, have thermal inertias lower than 300. Asteroids in other rotation bins exhibit a greater dependence of thermal inertia on skin depth than expected from \autoref{eq:skindepth}, which is possible when the porosity decreases for larger thermal inertia values. Therefore, this trend can be explained if these surfaces are covered with more boulders that have lower porosity than the surrounding regolith. It is also possible that this trend is a result of larger grain sizes, which we consider below.

As the relative contact radii between grains decreases for larger grains, the radiative heat transfer becomes dominant over solid heat transfer through grain contacts. In \citet[part I;][]{MacLennan+Emery21}, we showed that the variation in thermal inertias with temperature are consistent with the theoretical value when radiation is the dominant heat transfer mechanism. Additionally, both the \citet{Gundlach&Blum13} and \citet{Sakatani_etal18} models predict that the radiative thermal conductivity component is directly proportional to grain size. Therefore, we can expect that the thermal skin depth increases in proportion to the square-root of the grain size or, equivalently, that the grain size is proportional to the square of the skin depth. The lines showing this proportionality in the bottom panel of \autoref{eq:skindepth} shows that this is indeed the case across our entire dataset. Interestingly, it appears that objects with smaller rotation periods exhibit a stronger relationship between the predicted grain size and skin depth. Furthermore, objects with $d_g > l_s$ are more likely to be fast rotators as 10 out of 15 of these objects have $P_{rot} < 5 \hour$. In response to the question posed earlier in this section, we posit that the smaller skin depths caused by short rotation periods do indeed have a non-negligible effect on the thermal inertia and resulting regolith grain size. Yet, because most of the grain size estimates for these fast rotators are lower than the skin depth, we claim that this effect is minimal.

\subsection{Grain Size Factors Model}\label{subsec:MVR}

Multi-linear regression is a method that attempts to model a dependent variable (the grain size, $d_{g\phi}$) as a linear combination of several independent variables---in our case, diameter and rotation period. The fitted slope, or coefficient, of each independent variable and a $y$-intercept, along with 1$\sigma$ uncertainties for each of these parameters is reported. We consider several segmented, or piece-wise, multi-linear regression models with different numbers and combinations of break points for both independent variables ($D^b_\mathit{eff}$, and $P^b_\mathit{rot}$), as listed below:\begin{itemize}
\item {\it M-1}: No break-points
\item {\it M-2}: $1\times$\ $D^b_\mathit{eff}$
\item {\it M-3}: $1\times$\ $P^b_\mathit{rot}$
\item {\it M-4}: $1\times$\ $D^b_\mathit{eff}$, $1\times$\ $P^b_\mathit{rot}$
\item {\it M-5}: $2\times$\ $D^b_\mathit{eff}$
\end{itemize}
These segmented models partition the indicated independent variable into intervals for which a different function (slope) is fit to the data. Adding a break-point creates two lines that form a continuous, piece-wise function in lieu of a single linear fit. In addition to slope estimates of the lines, the location of the break-points are estimated.

The diameter and rotation period are transformed by taking the log$_{10}$ when used in the multi-linear regression models. This variable transformation is done to best capture the wide variance in these variables which each span more than 2 orders of magnitude. Unlike the thermal inertia multi-linear analysis presented in \cite{MacLennan+Emery21}, we do not consider temperature as an explanatory variable because it is already accounted for in the thermal conductivity model via temperature-dependent heat capacity values (\autoref{subsec:2.22}) and radiative heat transport.

We compare the multi-linear model fits to one another by taking the adjusted r-squared ($r^2_{adj}$) statistic and Bayesian Information Criterion ($BIC$). The $r^2_{adj}$ is a determination of the degree to which the model explains the variance in the dependent variable (i.e., higher values indicate a better fit), while also accounting for the number of predictor parameters, $w$, in the model. The number of free parameters is calculated from the \emph{total} number of fitted variables, which increases by 2 when a break-point is added \footnote{This is because each new line has a slope and a $y$-intercept that is independent from others, although a single model intercept is ultimately reported.} The adjusted r-squared is related to $r^2$ via: $r^2_{adj} = 1 - (1-r^2)(N-1)/(N-w-1)$.  The $BIC$ is used to indicate which model maximizes the likelihood of matching the data (lower values indicate a higher likelihood), and accounts for the number of model parameters (more parameters increases the score). It can be calculated via $BIC = N \ln{(RSS)} -N\ln{(N)} + w\ln{(N)}$ in which $RSS$ is the residual sum of squares \citep[see, e.g.][for more details]{Feigelson&Babu12}. Both of these statistics are shown in \autoref{tab:table2.4}, along with the number of model parameters.

When comparing the results amongst all models, we find that {\it M-4} has the largest $r^2_{adj}$ and lowest $BIC$. Statistically speaking, the difference between two model $BIC$ values, $\Delta BIC$, indicate a preference for one over another. \cite{Kass&Raferty95} state that $\Delta BIC > 6$ indicates a {\it strong} preference for the lower $BIC$. Since {\it M-4} has a $\Delta BIC = 34$ between it and the second lowest $BIC$ value ({\it M-2}), we use this as an indication that it is the preferred model. We plot the best-fit and preferred {\it M-4} model in \autoref{fig:modelfit}. Black symbols are estimated grain sizes for each object and colored symbols are values from the multi-linear models. Solid bars indicate the 1$\sigma$ range of the diameter and rotation period breakpoints, respectively, at the top of each panel. The estimated intercepts, linear coefficients (slopes), and break-point(s) ($D^b_\mathit{eff}$, $P^b_\mathit{rot}$), along with the associated uncertainties, are listed in \autoref{tab:table2.5}. We remark that the model intercept value represents the predicted $d_{g\phi}$ for a hypothetical object with $D_\mathit{eff} = 1 \km$ and $P_\mathit{rot} = 1 \hour$, which is just under $1\meter$ for {\it M-4}.

\subsection{Compositional Effects}\label{subsec:compgrainsize}

Here, we investigate if, and how much, the grain sizes vary by or depend on the surface composition of an asteroid. For this analysis we use spectral group as a proxy for composition. Simply comparing the means of the grain size distributions between the groups is not appropriate, as some of the independent explanatory variables are correlated with grain size: for example, primitive C-complex and P-type bodies are more represented at larger sizes, and the E/Xe-types are largely represented in the lower size range\footnote{Similar relationships are apparent between spectral classes and heliocentric distance \citep{DeMeo&Carry13}.}. Instead, we perform post-hoc t-tests between the spectral classes by using the model residuals from {\it M-4}. Since the multi-linear analysis doesn't account for the spectral type, any differences in the group residuals can be used to indicate disparities in regolith grain sizes.

We apply Welch's t-test\footnote{We use Welch's t-test, as opposed to the Student's t-test, as the latter assumes that the two groups have equal variance or sample size, which is not the case here.} \citep{Welch47} in a series of trials between each possible combination of spectral class, as well as between each class and the entire sample with that particular class removed. The null hypothesis that is tested is that the means between the groups do not differ. We report the $p$-values of these trials in \autoref{tab:compresid}: lower p-values indicate a higher probability that the null hypothesis is not supported. \autoref{tab:compresid} shows the mean and standard deviation of the model residuals and the number of objects for each group. We analyze the S, C, and E-types both with and without the objects in which their spectral type was inferred by their albedo. Similarly, for B-types we analyze both the confirmed B-types and those marked as B$^\star$ in \autoref{tab:table2.2} for which we assigned CM meteorite properties based on their low-thermal inertias. Consistent with their low thermal inertias, the B$^\star$ objects have consistently lower grain size residuals compared to confirmed B-types. This may indicate that these objects are not true B-types and may actually be closer in composition to P-types, which have the lowest grain sizes of all the spectral groups included here.

The model residual distributions, grouped by composition, along with the mean model residuals are depicted in \autoref{fig:residcomp}. Compared to the mean of the remainder of the sample, there is strong statistical evidence ($p<.01$) that suspected metal-rich asteroids on average exhibit grain sizes that are nearly twice as large as asteroids of the same diameter and rotation period. We interpret this difference as an indication that one of the thermophysical or material properties of FeNi metal affects the efficiency of regolith breakdown process. Additionally, the average grain sizes of M-types and E-types are statistically indistinguishable from one another (\autoref{tab:compresid}), and from the suspected metal-rich objects. On the other hand, P-types show a much smaller mean grain size than the remainder of the sample ($p < .01$). Finally, we note here that S-types and C-types have statistically smaller and larger grain sizes, on average, compared to the rest of the sample. Consistent with this finding, the carbonaceous K-types have coarser-grained regoliths compared to S-types. Potential explanations for these group differences in grain size are discussed in \autoref{sec:discussion}.

\subsection{Near-Earth Asteroids}\label{subsec:NEAanalysis}

Thus far in our analysis we have excluded NEAs and Mars-Crossers (MCs) in order to mitigate the potential influence on regolith evolution caused by widely-varying thermal and impact environments. Because these asteroids exhibit a wide range of orbital parameters (varying in both semimajor axis, $a$, and  eccentricity, $e$) they are subject to drastically varying external influence, whereas we wish to examine the factors inherent to asteroids themselves. We analyze the grain sizes (\autoref{grainsizeresults}) calculated for the NEAs with thermal inertia estimates from other works ($N = 21$) and those estimated in \citet{MacLennan+Emery21} ($N = 7$). With grain sizes for these 28 asteroids, we seek to identify regolith dependencies on orbital factors.

We perform a multi-linear regression model, similar to the one presented in \autoref{subsec:MVR}, but with different input factors. We do not consider segmented linear fits as the low number of objects in this NEA subset may result in model over-fitting. The independent explanatory variables we include here are the diameter and rotation period transformed into log$_{10}$ space. But now we also include the orbital semimajor axis ($a$), perihelion ($q$), and aphelion distance ($Q$) as possible explanatory variables. However, these orbital parameters are {\it not} transformed into logarithmic space in the multi-linear model and are left as-is.

The best-fit regression model for NEAs indicated that $Q$ is the only statistically significant variable. Thus, the NEA grain sizes dataset does not appear to show the same trends with the diameter and rotation period that exist among main-belt objects. On further inspection, potential co-linearity between $D_\mathit{eff}$ and $Q$ could raise some doubt of the significance of this result. However, employing a multi-linear model with only $Q$ and $D_\mathit{eff}$ did not change the significance of either variable. We thus conclude that $Q$ is a better predictor of grain size than diameter and rotation period for NEAs. The grain sizes of NEAs as a function of aphelion, perihelion, diameter, and rotation period are shown in \autoref{fig:NEAgrainsize}. Symbol colors indicate the spectral type of the objects, with unfilled symbols indicating that the classification was used using the albedo as a proxy. It is interesting to note here that the grain sizes of objects with $0.9 < q < 1.1 \au$ are somewhat higher (on average) than the asteroids outside this range and that asteroids with largest grain sizes ($> 10 \cm$) are found having perihelia interior to Earth's orbit.

An important caveat to note here is the fact that NEA and MC asteroid diameters extend to a lower size range compared to MBAs. Because the overlap in NEA and MBA diameters occurs for $2\km < D_\mathit{eff} < 40\km$, it is difficult to draw conclusions about the difference between the two populations from these sets of multi-linear models. Grain sizes for NEAs with $D_\mathit{eff} > 2~\km$ (\autoref{fig:NEAgrainsize}) are similar to those of similarly sized MBAs and exhibit the same inverse dependency on asteroid diameter. On the other hand, NEAs smaller than $2\km$ don't show any discernible trend between grain size and asteroid diameter. It is possible that MBAs exhibit a similar lack of trend, but the MBAs in our dataset do not extend to this size range. 

Additionally we note that, in general, the comparison between rotation period and grain size for NEAs appears to be consistent with MBAs, with a few exceptions. Asteroids with spin rates near the spin barrier ($P_\mathit{rot} \approx 2.12$ hr) exhibit a range of grain size spanning 3 orders of magnitude, which mimics the grain size span for the entire Main Belt sample. At the low end of the range is 1950~DA, with two other fast-rotating NEAs exhibiting very large grain sizes. The latter two appear to be very consistent with the inverse trend between rotation period and grain size for main-belt objects. Thus, 1950~DA appears to be an outlier in this respect and we further discuss this point below.

\section{Discussion}\label{sec:discussion}

The best-fit multi-variate model, {\it M-4}, indicates that a break-point value near $\sim10~\km$ occurs in the relationship between asteroid diameter and regolith grain size. The slope fit to objects smaller than this size shows a strong inverse relationship between asteroid diameter and regolith grain size. On average, the regolith grain size of 10~km bodies is 0.6~mm, with {\it M-4} predicting a grain size of 25~mm for 2~km objects---an increase in $d_{g\phi}$ of nearly a factor of fifty, which translates to a power slope of $\sim 6$. The {\it M-4} slope fit for for asteroids larger than $10~\km$ is statistically indistinguishable from zero, which suggests no dependence of regolith grain size on asteroid diameter. Interestingly, there appears to be an inverse relationship between the upper limit grain size and asteroid diameter among objects larger than $\sim 80 \km$. This may indicate a different process or an additional factor that is unaccounted for in our model that influences regolith development on 100-km-scale asteroids, which are through to be primordial bodies.

The preferred multi-linear model predicts that objects with $P_{rot} \approx 5\hours$, on average, have the smallest grain sizes when comparing across all rotation periods. As the rotation period increases, the model predicts that the grain size increases. Objects rotating faster than $\sim5\hours$ show a steep increase in grain size with decreasing rotation period. It is interesting to note that the model predicts that objects with $2.2\hours$ and $300\hours$ rotation periods should both have $d_g \sim~8\mm$. These grain size trends with rotation period could indicate at least one regolith evolution processes that depend on the spin rates of asteroids. We explore such processes in \autoref{subsec:fatigue} and \autoref{subsec:retentionloss}. Alternatively, the inverse trend at lower rotation periods could be a result of thermal skin depths that are comparable to the particle sizes, which we have shown this not to be the case in \autoref{subsec:skindepth}.

\subsection{Meteoroid Impact Breakdown}\label{subsec:breakdown}

Repeated impacts of smaller asteroids or meteoroids into the surface of an asteroid will create ejecta---some of which is retained at the surface as regolith. The general relationship between the mass of the ejecta ($M$) above a certain velocity ($u$), as a function of impactor mass and velocity ($m$ \& $U$) is:
\begin{equation}\label{eq:massvel}
	\frac{M}{m} \propto \Big(\frac{u}{U}\Big)^{-3\tau}
\end{equation}
where $\tau$, an empirically-derived exponent, is taken as 0.41---the value for sand \citep{Housen&Holsapple11}. For reference, the probability distribution of relative velocities between asteroids in the Main Belt has a mode of $4.3 \km \second^{-1}$ and a mean value of $5.3 \km \second^{-1}$ \citep{Farinella&Davis92,Bottke_etal94}. The velocities of incoming meteoroids are mostly independent of the size of target body (due to the extremely small gravitational attraction of an asteroid), which leaves the relative size of the impactor to the target asteroid as the dominant factor in \autoref{eq:massvel}.

Additionally, the mechanical and structural properties of the target asteroid play a role in determining the outcome \citep{Housen&Holsapple03}. For example, in targets with low porosity, the ejecta velocities are larger than that of an otherwise identical impact into a porous target \citep{Nakamura&Fujiwara91,Nakamura_etal94}. Energy from the impact is concentrated near the impact site and partitioned into crushing the surface material \citep{Flynn_etal15}, instead of ejecting it at high velocities. \cite{Holsapple_etal02} points out that if the porosity of the target region is greater than 50\%, only 10\% of the crater mass is ejected, as most of the energy is partitioned into compacting the material. This experimental finding was verified by \citet{Cambioni_etal21} who showed that the presence of fine-grained regolith was less likely to be found surrounding porous boulders on Bennu's surface. The authors claim that the production of fine-grained regolith in the general asteroid population is frustrated by the presence of porous boulders. If the surfaces of B-types contain porous boulders similar to Bennu's then we might expect a larger grain size compared to S-types, assuming that the latter are comprised of low porosity boulders. Inconsistent with this claim, the grain sizes of S-types and B-types are statistically similar for the asteroids in our sample. This most likely indicates that other factors may be more significant than porosity in determining the production of regolith. Further investigation into this topic should be pursued in future studies.

Using the measured meteorite flux at $1\au$, \cite{Basilevsky_etal13} calculated characteristic lunar boulder survival lifetimes from boulder size-frequency statistics on rims of lunar craters with known ages. They estimated that it takes 25--$50\Myr$ to destroy 50\% of a typical $1 \meter$-sized boulder, and after 190--$300\Myr$, more than 99 \% of boulders should be completely destroyed. \cite{Basilevsky_etal15} extrapolated the lunar timescale to Ceres and Vesta assuming that boulders are broken down exclusively via impact weathering. On these large MBAs the boulder weathering timescale is 3\% of the lunar value (i.e., 0.75--$1.5\Myr$), which is mainly a consequence of the meteorite flux being $\sim$2 orders of magnitude greater in the Main Belt than at $1~\au$. In all cases, the remaining fraction of boulders exponentially decreases with time.

This theoretical timescale is significantly shorter than the ages of most asteroid families, although a few young ($< 1 \Myr$) families have been identified \citep[e.g., Table 1 in ][]{Nesvorny_etal15} in the Main Belt. In principle, asteroids in these young families could theoretically be used to investigate regolith properties of asteroid surfaces that have been reset from an impact. One consideration that must be made is the fact that the small asteroids that exist in these recently-formed families are less likely to retain fine-grain regolith particles because of their low surface gravity. Statistically-speaking, a $1 \km$ asteroid is expected to experience an average of 5 catastrophic breakups every 1 Myr \citep{Holsapple_etal02}, or once every $\sim200 \kyr$. This timescale is shorter than the estimated boulder survival timescale (0.75--$1.5 \Myr$) of \citet{Basilevsky_etal15}, so we expect that an MBA of this size would not survive long enough to have much of a developed regolith. On the other hand, a $10 \km$ MBA will survive, on average, $16\Myr$ ($\sim 80$ times longer than a $1 \km$ body). The implication that $10 \km$ asteroids exist long enough to develop a regolith through impact degradation is consistent with our grain size results.

Our results show that the differences in grain size for NEAs and MCs are most likely dependent on their orbital parameters. This relationship could be explained by the drastic change in collisional probability and velocities of NEAs with MBAs in the Main-Belt \citep{Bottke_etal93}. Because an NEA that enter the Main-Belt is traveling relatively slower, the velocity difference between it and a typical MBA will be larger compared to two MBAs with similar orbits. This would theoretically result in more efficient impact weathering rate for NEAs with $Q \gtrsim 2.2 \au$, compared to MBAs of a similar size. However, there is no obvious difference in grain sizes for NEAs and MBAs of similar object sizes. Thus, we suppose that another process is the reason for the grain size dependency on orbital parameters.

Evidence from meteorite impact breccias---fragmented samples of an asteroid collision \citep{Burbine&Binzel02}---offers insight into to the state of asteroid regolith. These meteorites are formed from the lithification of near-surface material from the heat and pressure of an impact and classified by their texture and presence/absence of clasts \citep{Bischoff_etal06}. A direct connection to asteroid regolith can be made if they are rich in solar-wind gases that are identified isotopically \citep{McKay_etal89}. Clasts ranging in size from several hundred microns up to many centimeters have been found in HEDs, carbonaceous, ordinary, and enstatite chondrites, and other stony meteorite regolith breccias \citep[][and references within]{Bischoff_etal06}. Interestingly, \cite{Bischoff_etal06} point out that some aubrites are known to contain large enstatite clasts up to 10~cm in size, as well as metal grains up to 1~cm in size. If the aubrite parent body is matched by an E-type spectrum, then these systematically-larger clasts and grains are well-matched to the finding that E-types have larger regolith grains than other spectral types.

\subsection{Thermal Fatigue Breakdown}\label{subsec:fatigue}

The cyclic heating and cooling experienced by asteroid surfaces, as a consequence of time-varying insolation, can result in large spatial temperature gradients within the material. These gradients create heterogeneous thermal stress fields (expansion and contraction) across mineral grain boundaries of a rock \citep{Molaro_etal15}. This thermal fatigue process involves the structural weakening of the material from thermal gradients and ultimately results in the breakdown of asteroid regolith. It has been debated whether or not thermal fracturing is a relatively significant weathering mechanism for terrestrial rocks \citep[e.g.,][and references within]{Molaro&McKay10} yet \cite{Delbo_etal14} experimentally demonstrated the effectiveness of this process, in vacuo, on a chip of the Murchison (CM2 carbonaceous chondrite) and Sahara 97210 (L/LL 3.2 ordinary chondrite) meteorites. \citet{Libourel_etal21} performed a follow-up study with these two meteorites and Allende (CV3) where they demonstrated that larger temperature gradients, such as those experienced for low perihelion asteroids, result in greater crack growth rates. Images from the encounter of Bennu by OSIRIS-REx provided widespread evidence of thermal cracking across the surface \citep{Molaro_etal20}. Convincing evidence from the Moon and Mars \citep{Ruesch_etal20,Eppes_etal15} indicates that thermal breakdown is efficient on other solar system bodies. Both \citet{Molaro_etal17} and \citet{Ravaji_etal19} posit that the efficiency of thermal fracturing is controlled by the maximum thermal stress experienced by the rock, as opposed to the average, so we focus on this quantity in our discussion.

\cite{Molaro&Byrne12} numerically simulated the amount of internal stress experienced by a rock on Vesta, Mercury, and the Moon's surface. In particular, they estimated the amount of temperature change, per unit time ($\sfrac{\Delta T}{\Delta t}$) and compared to the temperature gradient $\nabla$T within rocks having orientations at different latitudes on the surface. They found that while a rock on Vesta experienced the largest temporal temperature change, it also had the lowest $\nabla$T---leading to a smaller stress field. This case is opposite to that of the Moon and Mercury, in which $\nabla$T and $\sfrac{\Delta T}{\Delta t}$ were highly correlated. The modeling efforts of \cite{Molaro&Byrne12} demonstrated that greater thermal stresses were experienced for rocks experiencing faster sunrises (i.e., shorter rotation periods). The stresses were more pronounced for rock surfaces that faced the rising sun. Daytime shadowing, especially when occurring just after local sunrise or before local sunset, was also a major contributor to increasing the temperature gradients. The authors also predict that pre-existing cracks would contribute even more to the crack growth rate and shorten breakdown timescales.

Our asteroid grain size modeling results indicate that objects with $P_{rot} = 5\hours$ are most likely to exhibit the fine-grained regoliths and coarse-grained regoliths ($d_g \gtrsim 1\cm$) are found among the asteroids with $P_\mathit{rot} > 100 \hours$. This result coincides with the findings of \cite{Molaro_etal17}, \citet{ElMir_etal19}, and \citet{Ravaji_etal19} as they all predict less-efficient thermal fatigue for objects with longer rotation periods. The rotation period breakpoint in the multilinear fit at $\sim5\hours$ could be indicative of an optimal rotation rate that maximizes thermal fracturing for asteroid surfaces. The \cite{Molaro_etal17} model do not predict any rotation period that is optimized thermal fracturing efficiency, yet \citet{ElMir_etal19} and \citet{Ravaji_etal19} found that breakdown timescales for a $10\cm$ rock are minimized for rotation periods in the range $10-15\hours$. It is not immediately obvious why these works predict a slightly larger optimal rotation rate for thermal breakdown. Future works should seek to reconcile thermal fatigue models with our empirical findings.


\cite{Molaro_etal15} studied the sensitivity of their micro-stress model to changes in the material properties of the rock and found the thermal expansion coefficient and Young's modulus (which describes tensile elasticity) to be the most influential properties, and that thermal conductivity, perhaps counter-intuitively, does not significantly affect the rate of crack growth. The thermal parameter,
\begin{equation}
    \Theta = \frac{\Gamma}{\varepsilon \sigma_0 T_{eq}^3} \sqrt{\frac{2 \pi}{P_\mathit{rot}}}
\end{equation}
may intuitively at first seem to be useful in predicting the effectiveness of thermal fatigue on asteroid surfaces, as it can be used as a proxy for the amplitude of the diurnal temperature range. However, its formulation doesn't capture the essential information regarding the heating and cooling {\it rates} of a boulder/rock at the surface. For example, regolith on very slow rotators will experience slow sunrises that result in relatively slow rate of temperature change. This behavior may seem paradoxical if one were only to examine the thermal parameter for a hypothetical slow-rotator: as $P_{rot}$ increases, $\Theta$ becomes smaller and the diurnal temperature range is maximized. However, the rate of temperature change is relatively small for a hypothetical point on the surface of a slow-rotator, and the insolation changes experienced during long sunrises and sunsets do not result in spatial temperature gradients large enough to cause significant thermal fracturing.

Our results indicate that the grain size of E-type (including assumed E-types), M-type, and suspected metal-rich (Met) asteroids are, on average, $\sim2$ times higher than asteroids of the same diameter and rotation period (\autoref{fig:modelfit}). This suggests that either regolith generation is inefficient for these bodies or that smaller regolith grains are preferentially lost (see \autoref{subsec:retentionloss}). Considering the former, we consider the idea that compositional differences may lead to less efficient breakdown. For example, the relatively high values of Young's moduli for FeNi metal should result in more efficient breakdown via thermal fatigue \citep{Molaro_etal15} because stiffer materials are less able to accommodate the differential thermal expansion of grains. This idea is supported if we consider the heterogeneous mesosiderite meteorites, which consist of a silicate-metal mixture, as the appropriate analog for metal-rich bodies. We suspect that greater thermal stresses surrounding silicate grains will lead to efficient thermal fatigue breakdown. Therefore we suspect that E-types, M-types, and metal-rich bodies are subject to one or more processes that removes smaller particles from their surfaces.

Finally, we remark that thermal fatigue may be a self-limiting process, as long as other processes are insignificant. In this case we claim that an insulating layer of regolith may form over a boulder as a result of thermal fatigue, effectively shielding it from larger temperature gradients. In actuality, impact processes disrupt asteroid surfaces over time and overturn the topmost layers \citep[i.e., impact gardening;][]{Horz77}. This may lead to regolith stratification over long periods of time, depending on the impact gardening timescale and efficiency relative to thermal breakdown. Therefore, asteroid regolith may become stratified at mm or sub-mm length scales with progressively smaller grains near the surface.

\subsection{Regolith Retention \& Loss}\label{subsec:retentionloss}

The discovery and characterization of active asteroids---those which are observed to exhibit mass-loss---is direct evidence that asteroids are subject to processes that remove regolith from their surface. Several potential mechanisms have been proposed and some remain as only potential explanations the activity. \citet{Jewitt12} reviews such mechanisms, which include impact, centrifugal (inertial), and electrostatic ejection. The OSIRIS-REx cameras directly observed particle ejection from Bennu's surface, which \citet{Lauretta_etal19} posited were caused by thermal fracturing, volatile dehydration, or meteoroid impacts. Other plausible particle ejection mechanisms such as electrostatic charging were ruled out for Bennu but may act on other objects' surfaces. Depending on the physical properties of an asteroid and its orbital location, all of these processes may or may not be relevant or efficient. We also note that each mechanism may alter asteroid surfaces in different ways because their effectiveness is dependent on the grain size.

In addition to generating regolith, meteoroid impacts may also \emph{remove} regolith grains from an asteroid surface. As noted above, the energy from a meteoroid impact upon an asteroid surface is partly transferred to individual regolith grains, resulting in a velocity distribution that varies with grain size \citep{Takasawa_etal11}. As remarked by \citet{Lauretta_etal19}, high energy impacts into asteroid surfaces (large impactor mass and/or velocity) do not produce ejecta, but instead partition their energy into modifying the target material \citep[e.g.][]{Fiege_etal19}. The smallest regolith particles have higher velocities and are thus the most likely to be ejected. The upper size limit for particle ejection depends on factors such as the impactor mass and target mass (\autoref{eq:massvel}). These variables can take on a wide range of values varied in order to predict the mass fraction of ejecta that is retained as a function of velocity. The active asteroids (596) Scheila ($D_\mathit{eff} \sim 113\km$) and P/2010 R2 (La Sagra) ($D_\mathit{eff} \sim 1.4\km$) are suspected to have been subject to impact-induced mass loss due to their single and sudden brightening events \citep{Jewitt12}.

The material strength of the target plays a vital role in the ejecta outcome. In general, stronger surface materials produce more ejecta, which can escape the gravitational well of the body more easily than an asteroid comprised of weaker material. In addition, \citet{Matsui&Schultz84} show that impacts into brittle (low temperature) metal will result in fracture and spallation compared to ductile metal at higher temperature. For high energy impacts, the strength and/or porosity of the object can play a similar role as the surface material does for low energy impacts. We expect that $\sim100\km$ objects that exhibit larger regolith grains could be stronger and/or less porous than those with smaller regolith grains. This could explain why the suspected metal-rich asteroids in our sample, all of which are $> 10\km$, have larger than average grain sizes. Future efforts in mass determination for many asteroids can be used to investigate a potential relationship between the macro-porosity of asteroids and regolith development.

Our results show that, on average, grain sizes for asteroids larger than $\sim10 \km$ don't show any dependence on object size. On the other hand, the average grain size increases with decreasing asteroid size for objects smaller than $10\km$. This size cutoff could be indicative of an abrupt change in the process of regolith breakdown, or alternatively, may indicate a change in object physical properties (i.e., interior strength) that controls the mass of impact ejecta. Using crater scaling laws, \cite{Housen_etal79} predict that asteroids in the 1--10 km range should not harbor a significant regolith ($\ll 1 \mm$ thick) due to a decrease in \emph{both} the strength and gravitational field of bodies of this size. Asteroids with higher strengths have an increased ejection velocities, for similar impact scenarios, compared to low strength objects \citep{Nakamura_etal94}. Strength could be coincident with the material composition, which could explain why metal-rich bodies, which are presumably stronger, have less fine-grained regolith compared to S-complex and C-complex asteroids.

For fast-rotators, the outward centripetal acceleration near their equators is comparable to the downward gravitational acceleration. Regolith grains in these locations therefore exist in a precarious state, and a small perturbation can transfer enough energy to cause the ejection of equatorial material where the effective gravity is near-zero \citep{Guibout&Scheeres03}. It is difficult to theoretically predict whether this kind of mass-loss occurs incrementally, on a grain-by-grain basis, or catastrophically, with large portions ejected at a time \citep{Scheeres15}. \citet{Lauretta_etal19} ruled out centrifugal ejection as an ejection mechanism for Bennu and, to date, centrifugal ejection has not been definitively determined for any active asteroid \citep{Jewitt12}.

Among small asteroids, the cohesive (e.g., van der Waals) forces between regolith grains smaller than $\sim 1\cm$ have theoretically \citep{Scheeres_etal10} and experimentally \citep{Murdoch_etal15} been shown to dominate over gravitational, inertial, and, in some cases, electrostatic forces. Evidence for cohesive forces present on an asteroid surface was found for the $1.3 \km$ NEA (29075) 1950~DA \citep{Rozitis_etal14,Gundlach&Blum15}. Cohesive forces are inversely dependent on the surface area of the regolith grains \citep{Scheeres_etal10}. This suggests that a regolith comprised of smaller grains may be stronger than one comprised of larger grains, which has implications for the effectiveness and behavior of potential mass-loss mechanisms. Particles or rocks with low cohesion could be individually ejected, yet inter-particle forces can cause the regolith to be more susceptible to large-scale structural failure. In this scenario planes of weakness preferentially form around massive clumps of grains \citep{Sanchez&Scheeres20}. As these grain structures are held in a higher energy state, even a small perturbing force that exceeds the yield stress could trigger a landslide \citep{Scheeres_etal10}, potentially resulting in a catastrophic loss of fine-grained regolith or, more likely, a mixture of regolith and boulders.

The best-fit multilinear model predicts that faster rotators (with $P_{rot} < 5\hours$) have larger regolith grains. Generally speaking, this finding is consistent with thermal breakdown of regolith and/or with centrifugal ejection of smaller grains. Because smaller grains have greater cohesive properties, we slightly favor the former, although it is possible that centrifugal mass loss can affect a wide range of particle sizes.  Interestingly, two fast rotators in the sample, (29075) 1950~DA ($D_\mathit{eff} \approx 1.3 \km$, $P_\mathit{rot} \approx 2.12$) and (3554) Amun ($D_\mathit{eff} \approx 2.7\km$, $P_\mathit{rot} \approx 2.53 \hours$), posses regolith characteristics at the extremes of the sample: Amun has a high thermal inertia, suggesting it may be completely devoid of fine-grained regolith, in contrast to 1950~DA \citep{Rozitis_etal14}. Amun may have always been lacking any significant regolith or it could had been similar to 1950~DA in the past. In the latter case, it was subsequently and periodically spun-up such that centrifugal forces exceeded any cohesion holding the regolith together resulting in the loss of fine-grained regolith held together in massive clumps, as described above. In this scenario, we might expect 1950~DA to lose clumps of fine grained material as its rotation rate incrementally increases and structural failure occurs within the regolith. More thermal inertia estimates of rapid rotators ($P_{rot} < 3 \hours$) may provide greater detail into the efficiency and probability of centrifugal ejection and the cohesive strength of regolith \citep{Rozitis_etal14}.

Regolith particles can become electrostatically charged when exposed to ionized solar wind particles, which are primarily comprised of protons \citep{Lee96}. Solar UV radiation releases photoelectrons from the surface and a negatively-charged ``sheath'' builds up on the sunlit side of an asteroid. The electrostatic forces that regolith grains experience in this field can {\it potentially} overcome the cohesive and gravitational forces \citep{Hartzell19}. For smaller grains, cohesive forces will always dominate and electrostatic forces are not likely to remove them from the surface. The largest grains, on the other hand, are too heavy to be affected by electrostatic forces. However, there is an optimal size---proportional to $g_a^{1/4}$---for which electrostatic forces overcome cohesive and gravitational forces, resulting in grain levitation \citep{Hartzell&Scheeres11}. The optimal grain size is estimated by \cite{Hartzell&Scheeres11} to be around 2--$4~\cm$ for an Eros-sized object ($g_a \sim 5 \times 10^{-3}\meter \second^{-2}$) and 10--$30~\cm$ for an Itokawa-sized object ($g_a \sim 8 \times 10^{-5} \meter \second^{-2}$).

If electrostatic forces dominate over cohesive forces, we would then expect a {\it positive} correlation between asteroid size and regolith grain size as opposed to an inverse relationship. Although the entire NEA grainsize dataset doesn't show any correlation with asteroid size, the sub-km NEAs in our dataset do exhibit a positive correlation between effective asteroid diameter and grain size that is consistent with this expectation. We suspect that regolith processes in this size range become confounded by multiple competing forces and future modeling could show which process(es) dominate on different bodies.

\section{Conclusions and Future Work}\label{conclusion}

In this work we have presented regolith grain size estimates for 452 asteroids (\autoref{grainsizeresults}) and performed a few multi-linear models fits to the dataset and {\it post-hoc} tests between compositional groups and asteroid families. From our results we conclude the following:

\begin{quote}
\begin{enumerate}

\item[1)] Regolith grain size across our sample of main-belt asteroids is inversely dependent on object size when the effective diameter is below $\sim10\km$ (\autoref{subsec:MVR}). We suspect that this relationship is most likely due to the removal of smaller regolith particles via impact processes due to the lower surface gravity of smaller objects. For MBAs larger than $\sim10\km$, regolith grain size does not exhibit any dependence on asteroid size. Additionally, grain size shows a positive dependence on rotation period for $P_{rot}> 5\hours$ and an inverse relationship below this value. Nearly all the asteroids with rotation periods above $100\hours$ are coarse-grained with grain sizes $>1\cm$.

\item[2)] We show that potential metal-rich asteroids have consistently larger regolith grains than objects belonging to other spectral types of similar size and rotation period \autoref{subsec:compgrainsize}. We find that M-types and E-types have coarser-grained regoliths, which may be related to their enstatite content. Finally, the compositionally-primitive P-types exhibit lower-than-average regolith grain sizes (more fine-grained) than any other spectral type.


\item[3)] The regolith grain sizes of NEAs, which represent asteroid diameters smaller than the MBAs in our sample, are inversely dependent on aphelion (\autoref{subsec:NEAanalysis}) but not diameter or rotation period. Additionally, the grain sizes of NEAs with aphelia in the Main-Belt are consistent with MBAs with similar aphelia, and the grain sizes of $>2\km$ NEAs are consistent with that of MBAs at the same sizes. Thus, it is not possible to say if NEAs have different regoliths than MBAs, or if aphelion, rather than size, is a controlling factor of NEA grain sizes.

\item[4)] Finally, we infer from our analyses that evidence for both impact weathering and thermal fatigue/fracture exists in the asteroid population. In future work, processes that remove regolith (\autoref{subsec:retentionloss}), which may preferentially act on smallest regolith particles or largest boulders, should be accounted for when modeling asteroid regolith evolution.

\end{enumerate}
\end{quote}

Although the thermal conductivity equations in this work are useful to estimate grain size from thermal inertia, more effort is needed to improve modeling efforts to better resemble asteroid surfaces and regoliths. In particular, considerations for mixed surfaces comprised of heterogeneous grains, and both solid and porous boulders should be sought \citep[e.g.][]{Cambioni_etal19}.

Future thermophysical investigations of asteroids should target objects belonging to categories for which there are a relatively low number of thermal inertia estimates. For example, MBAs smaller than $\approx 3 \km$, super slow rotators ($P_\mathit{rot} > 100 \hours$), and members of very young asteroid families ($<10 \Myr$). In addition, thermophysical investigation of the regoliths of object groups not studied in this work (e.g., contact binary asteroids, asteroids with satellites, objects with close planetary encounters, low perihelion asteroids, etc.) would further elucidate regolith evolution processes and their respective timescales.

\section*{Acknowledgements}

We appreciate the comments and critiques from two anonymous reviewers that enhanced the quality and efficacy of this article. This work is made possible by funding from the NASA Earth and Space Science Fellowship NNX14AP21H. E.M.M. is partly supported by the Academy of Finland (\#316292). Article open access funded by Helsinki University Library.





\bibliographystyle{apalike}
\bibliography{references}







\begin{table}[p!]
\caption{Meteorite Thermal Conductivity Measurements at 200 K}
\label{tab:porosityk}
\scriptsize
\begin{center}
\begin{tabular}[h!]{llSSr}

	\hline \vspace{-0.2cm} \\
	Sample & Group & $\phi$\hyperlink{TMa}{$^a$} & $k_\mathit{eff}$\hyperlink{TMb}{$^b$} & \multicolumn{1}{c}{Source} \vspace{0.2cm} \\ \hline \\[-1,4em] \hline \vspace{-0.2cm} \\
	ALH 77288 & H6 & 2.0 & 3.53 & {\cite{Yomogida&Matsui83}} \\
	Arapahoe & L5 & 2.5 & 2.31 & {\cite{Yomogida&Matsui83}} \\
	Bath Furnace 1 & L6 & 4.3 & 2.26 & \cite{Opeil_etal12} \\
	Bath Furnace 2 & L6 & 4.3 & 2.72 & \cite{Opeil_etal12} \\
\end{tabular}
\end{center}
\end{table}

\begin{table}[pb!]
\caption{Meteorite Heat Capacity Measurements at Different Ambient Temperatures}
\label{tab:heatcapT}
\begin{center}
\scriptsize
\begin{tabular}[h]{llccr}

	\hline \vspace{-0.2cm} \\
	Sample & Type & $T_\textrm{lab}$ (K) & $c_s$ (J kg$^{\textrm{-}1}$ K$^{\textrm{-}1}$) & \multicolumn{1}{c}{Source} \vspace{0.2cm} \\ \hline \\[-1,4em] \hline \vspace{-0.2cm} \\
	Abee & EH4 & 200 & 500 $\pm$ \enskip  10 & {\cite{Yomogida&Matsui83}} \\
	Allende & CV & 180 & 501 $\pm$ \enskip 21 & \cite{Consolmagno_etal13} \\
	Bilanga & Di. & 180 & 509 $\pm$ \enskip 26 & \cite{Consolmagno_etal13} \\
	Bori & L6 & 180 & 495 $\pm$ \enskip \enskip 8 & \cite{Consolmagno_etal13} \\
\end{tabular}
\end{center}
\end{table}

\afterpage{
\begin{landscape}
\thispagestyle{mylandscape}
\begin{small}

\begingroup
\setlength{\LTleft}{-40cm plus -1fill}
\setlength{\LTright}{\LTleft}
\begin{longtable}{llcccccc}

\multicolumn{8}{l}{{\bf Notes. }Choices for the thermal conductivity, heat capacity, and emissivity values are explained in \autoref{subsec:2.22}. {\it Italicized} values are taken from S-complex values.}\\ \multicolumn{7}{l}{Low-$k$ refers to weak, low conductivity material that is not represented in meteorite collections.} \\
\multicolumn{7}{l}{\hypertarget{T5a}{$^a$}{All calculated specific heat capacities are assigned a 10\% uncertainty.}} \\
\multicolumn{7}{l}{\hypertarget{T5b}{$^b$}{Poisson's ratio.}} \\
\multicolumn{7}{l}{\hypertarget{T5c}{$^c$}{Average using data from: \cite{Consolmagno_etal09,Opeil_etal10,Szurgot_etal12}.}} \\
\multicolumn{7}{l}{\hypertarget{T5d}{$^d$}{Calculated from $E$ and $G$ values presented in \cite{Ibrahim12}.}} \\
\multicolumn{7}{l}{\hypertarget{T5e}{$^e$}{\cite{Macke_etal11a}}} \\
\multicolumn{7}{l}{\hypertarget{T5f}{$^f$}{\cite{Macke_etal11b}}} \\
\multicolumn{7}{l}{\hypertarget{T5g}{$^g$}{\cite{Macke_etal10}}} \\
\multicolumn{7}{l}{\hypertarget{T5h}{$^h$}{\cite{Macke10}}} \\
\multicolumn{7}{l}{\hypertarget{T5i}{$^i$}{\cite{Ibrahim12}}} \\
\multicolumn{7}{l}{\hypertarget{T5j}{$^j$}{\cite{Ledbetter&Reed73}}}
\endfoot

\caption{{\normalsize Material Properties of Meteorite Groups}}\label{tab:table2.2}\\

	\hline \vspace{-0.2cm} \\
	Spectral Group & Material Analog & $k_\mathit{grain}$ (W m$^{-1}$ K$^{-1}$) & $\rho_\mathit{grain}$ (kg m$^{-3}$) & $c_s$\hyperlink{T5a}{$^a$} (J kg$^{-1}$ K$^{-1}$) & $\epsilon_g$ & $E$ (GPa) & $\nu$\hyperlink{T5b}{$^b$} \vspace{0.3cm} \\ \hline \\[-1,6em] \hline \vspace{-0.2cm} \\
	S-complex & ordinary chondrites & 4.05 $\pm$ 0.35 & 3180--3710\hyperlink{T5h}{$^h$} & -0.0033$\times$T$^2$ + 3.39$\times$T & 0.9 & 28.8 $\pm$ 2.4\hyperlink{T5i}{$^i$} & 0.23 $\pm$ 0.04\hyperlink{T5d}{$^d$} \\
	V-type & HEDs & 4.05 $\pm$ 0.35 & 3180--3440\hyperlink{T5e}{$^e$} & -0.0033$\times$T$^2$ + 3.39$\times$T & 0.9 & 28.8 $\pm$ 2.4\hyperlink{T5i}{$^i$} & 0.23 $\pm$ 0.04\hyperlink{T5d}{$^d$}\vspace{0.2cm} \\  \hline \vspace{-0.3cm} \\
\end{longtable}
\endgroup
\end{small}
\end{landscape}
}

{\begin{landscape}
\pagestyle{mylandscape}
\newpage
\renewcommand{\arraystretch}{1.2}
\begin{scriptsize}
\begingroup
\setlength{\LTleft}{-50cm plus -1fill}
\setlength{\LTright}{\LTleft}
\begin{longtable}{lcccccccccccr}

\caption{{\normalsize Grain Size Model Inputs and Results}}\label{tab:table2.3} \\

 \hline \vspace{-0.3cm} \\
 Object & $D_\mathit{eff}$ & $p_V$ & $\Gamma$\hyperlink{T6a}{$^a$} & Source & \multicolumn{1}{c}{Tax/Color} & $P_\mathit{rot}$ & \multicolumn{1}{c}{$T_\mathit{surf}$ (K)} & Model Comp. & $d^{G\&B}_{g\phi}$ & $d^{Sak}_{g\phi}$ & $l_s$ (mm) & \multicolumn{1}{c}{Family} \vspace{0.2cm} \\ \hline \\[-1,6em] \hline \vspace{-0.2cm} \endfirsthead

 \multicolumn{13}{c}{{\normalsize \tablename\ \thetable{} --- continued}} \vspace{0.1cm} \\
 \hline \vspace{-0.3cm} \\
 Object & $D_\mathit{eff}$ & $p_V$ & $\Gamma$\hyperlink{T6a}{$^a$} & Source & \multicolumn{1}{c}{Tax/Color} & $P_\mathit{rot}$ & \multicolumn{1}{c}{$T_\mathit{surf}$ (K)} & Model Comp. & $d^{G\&B}_{g\phi}$ & $d^{Sak}_{g\phi}$ & $l_s$ (mm) & \multicolumn{1}{c}{Family} \vspace{0.2cm} \\ \hline \\[-1,6em] \hline \vspace{-0.2cm} \endhead

 \vspace{-0.3cm} \\ \hline
 \endfoot

 \vspace{-0.3cm} \\ \hline
 \multicolumn{13}{l}{{\bf Notes \& References.}
 \hypertarget{T1}{$^1$}SMASS I; \hypertarget{T2}{$^2$}SMASSS II; \hypertarget{T3}{$^3$}S3OS2 \citep{Lazzaro_etal04}; \hypertarget{T4}{$^4$}MITHNEOS; \hypertarget{T4b}{$^{4b}$}\cite{Tholen84}; \hypertarget{T5}{$^5$}\cite{GietzenLacy07}; \hypertarget{T6}{$^6$}\cite{Reddy_etal12b}; \hypertarget{T7}{$^7$}\cite{JewittHsieh06}} \\
 \multicolumn{13}{l}{\hypertarget{T8}{$^8$}\cite{ChapmanGaffey79}; \hypertarget{T9}{$^9$}\cite{Hardersen_etal11}; \hypertarget{T10}{$^{10}$}\cite{Shepard_etal06}; \hypertarget{T11}{$^{11}$}\cite{Vilas_etal98}; \hypertarget{T12}{$^{12}$}\cite{FeiberBeyer15}; \hypertarget{T13}{$^{13}$}MacLennan et al., (in prep)} \\
 \multicolumn{13}{l}{\hypertarget{TI1}{[T1]}\cite{AliLagoa_etal20}; \hypertarget{TI2}{[T2]}\cite{Capria_etal14}; \hypertarget{TI3}{[T3]}\cite{Marsset_etal17}; \hypertarget{TI4}{[T4]}\cite{Matter_etal13}; \hypertarget{TI5}{[T5]}\cite{Keihm_etal12}; \hypertarget{TI6}{[T6]}\cite{Delbo_etal15}; \hypertarget{TI7}{[T7]}\cite{Jiang_etal20};} \\
 \multicolumn{13}{l}{\hypertarget{TI8}{[T8]}\cite{Muller+Blommaert04}; \hypertarget{TI9}{[T9]}\cite{Hanus_etal18}; \hypertarget{TI10}{[T10]}\cite{MacLennan+Emery21}; \hypertarget{TI11}{[T11]}\cite{Marciniak_etal19}; \hypertarget{TI12}{[T12]}\cite{Marciniak_etal18}; \hypertarget{TI13}{[T13]}\cite{MacLennan+Emery19};} \\
 \multicolumn{13}{l}{\hypertarget{TI14}{[T14]}\cite{Yu_etal17}; \hypertarget{TI15}{[T15]}\cite{Hanus_etal15}; \hypertarget{TI16}{[T16]}\cite{Rozitis_etal18}; \hypertarget{TI17}{[T17]}\cite{Pravec_etal19}; \hypertarget{TI18}{[T18]}\cite{Rozitis_etal13}; \hypertarget{TI19}{[T19]}\cite{Leyrat_etal11};} \\
 \multicolumn{13}{l}{\hypertarget{TI20}{[T20]}\cite{Hanus_etal16}; \hypertarget{TI21}{[T21]}\cite{Rozitis_etal14}; \hypertarget{TI22}{[T22]}\cite{Devogele_etal20}; \hypertarget{TI23}{[T23]}\cite{Muller_etal04}}\\
 
 \multicolumn{13}{l}{\hypertarget{TTdag}{$^\dagger$}Indicates that the reported $P_\mathit{rot}$ value has been rounded to four decimal places.} \\
 \multicolumn{13}{l}{{$^\star$t} The reported grain size is larger than the estimated thermal skin depth.} \\
 \multicolumn{10}{l}{\hypertarget{T6a}{$^a$}{Thermal inertia values are in SI units (J m$^{-2}$K$^{-1}$s$^{-1/2}$).}} \\
 \multicolumn{10}{l}{\hypertarget{T6c}{$^C$}{Colors from \citep{Tholen84} are used.}} \\
 \multicolumn{10}{l}{\hypertarget{T6e}{$^e$}{The mutual eclipse duration of 4 hr is used to compute thermal skin depth.}} \\
 \multicolumn{10}{l}{\hypertarget{T6f}{$^F$}{Family association (\autoref{tab:table2.7}) is used.}} \\
 \multicolumn{10}{l}{\hypertarget{T6o}{$^O$}{Family outlier, with suspected family crossed out in last column. See text for details.}} \\
 \multicolumn{10}{l}{\hypertarget{T6p}{$^A$}{Object albedo (\autoref{fig:albG}) is used to infer meteorite connection.}} \\
 \multicolumn{10}{l}{\hypertarget{T6s}{$^S$}{Reflectance spectrum is used. See above for sources.}}\\
 \multicolumn{10}{l}{\hypertarget{T6t}{$^T$}{Large thermal inertia is used to infer a metal-rich surface.}}\\
 \multicolumn{10}{l}{\hypertarget{T6u}{$^U$}{A uniform, rather than a Gaussian, distribution was used for the thermal inertia.}}\\
 \endlastfoot

\input{grainresults_finfix2.txt}
\end{longtable}
\endgroup
\end{scriptsize}
\end{landscape}
}

\begin{table}[h!]
\caption{Comparison of Multi-linear Regression Models}
\label{tab:table2.4}
\begin{center}
\begin{tabular}[pt]{llcc}

	\hline \vspace{-0.2cm} \\
	Model & $w$ & $r^2_{adj}$ &$BIC$ \vspace{0.3cm} \\ \hline \\[-1,6em] \hline \vspace{-0.2cm} \\
	{\it M-1} & 4 & 0.062 & 1687 \\
	{\it M-2} & 6 & 0.141 & 1660 \\
	{\it M-3} & 6 & 0.149 & 1656 \\
	{\it M-4} & 8 & 0.217 & 1617 \\
	{\it M-5} & 8 & 0.141 & 1670\vspace{0.2cm}\\ \hline

\end{tabular}
\end{center}
\end{table}

\begin{table}[h!]
\caption{Linear Coefficients and Intercept for {\it M-4}}
\label{tab:table2.5}
\begin{center}
\begin{tabular}[t]{l|ccc}

	 \hline \vspace{-0.2cm} \\
	 Parameter & {Estimate} & 1$\sigma$ Uncertainty & {$p$-value} \vspace{0.3cm} \\ \hline \\[-1,6em] \hline \vspace{-0.2cm} \\
	 intercept [$d_{g\phi}$] & -9.85 & $\pm$ 1.27 & \cellcolor{Gray}$<$ .01 \\
	 $D^b$ [km] & 10.1 & +1.2/-1.1 & \cellcolor{Gray}$<$ .01 \\
	 $P^b$ [hr] & 4.84 & +0.35/-0.32 & \cellcolor{Gray}$<$ .01 \\
	  $D^\textrm{slope}_\mathit{eff}\ (<D^b)$ & 5.27 & $\pm$ 0.89 & \cellcolor{Gray}$<$ .01 \\
	 $D^\textrm{slope}_\mathit{eff}\ (>D^b)$ & -0.42 & $\pm$ 0.24 & \space \ \ \ .11 \\
	 $P^\textrm{slope}_{rot}\ (<P^b)$ & 7.83 & $\pm$ 1.70 & \cellcolor{Gray}$<$ .01 \\
	 $P^\textrm{slope}_{rot}\ (>P^b)$ & -2.04 & $\pm$ 0.28 & \cellcolor{Gray}$<$ .01\vspace{0.2cm}\\\hline

\end{tabular}
\end{center}
\end{table}

\afterpage{
\footnotesize
\begin{longtable}{lrrlcl}
\caption{\normalsize Asteroid Family Properties}\label{tab:table2.7}\\

\hline \vspace{-0.3cm} \\
PDS Family ID & AstDyS Family ID & Num & \multicolumn{1}{c}{Spec.} & Collision Type\hyperlink{Tf1}{$^1$}/Note & \multicolumn{1}{c}{Count\hyperlink{Tf2}{$^2$}}
\vspace{0.3cm} \\ \hline \\[-1,6em] \hline \vspace{-0.2cm} \endfirsthead

\multicolumn{5}{c}{{\normalsize \tablename\ \thetable{}---continued}} \\ \vspace{-0.2cm} \\
	\hline \vspace{-0.2cm} \endhead

\vspace{-0.2cm} \\ \hline
\endfoot

\vspace{-0.2cm} \\ \hline \vspace{-0.3cm} \\
\multicolumn{5}{l}{\bf Notes.} \\
\multicolumn{4}{l}{\hypertarget{Tfs}{$^*$}Largest family member appears in this study.} \\
\multicolumn{4}{l}{\hypertarget{Tf1}{$^1$}Collisional type definitions from \cite{Milani_etal15}.} \\
\multicolumn{4}{l}{\hypertarget{Tf2}{$^2$}Number of asteroid family members present in this study.} \\

\endlastfoot
    003 Hungaria & (434) Hungaria & 2965 & Xe/E & Fragmentation & 6 \\
    \multicolumn{6}{c}{\it Inner Main Belt} \\
    401 Vesta & (4) Vesta & 15252 & V & Cratering & 14\hyperlink{Tfs}{$^\star$} \\
    402 Flora & \multicolumn{1}{c}{|} & 13786 & S & & 25\hyperlink{Tfs}{$^\star$} \\
    403 Baptistina & (883) Matterania & 2500 & S & cluster within Flora & 3\hyperlink{Tfs}{$^\star$} \\
    404 Massalia & (20) Massalia & 6424 & S & Cratering & 1\hyperlink{Tfs}{$^\star$} \\
    405 Nysa-Polana high-pV) & & & S & & 8 \\
    405 Nysa-Polana (low-pV) & & & C & & 3 \\
\end{longtable}
}

\begin{table}[p!]
\caption{Residual grain size statistics and Welch's t-test results for different spectral groups}\label{tab:compresid}
\footnotesize
\begin{center}
\begin{tabular}[h]{lr|ccccccccc|cccc}
	\hline \vspace{-0.3cm} \\
	& & \multicolumn{9}{c|}{Spectral Group} & \multicolumn{4}{c}{Assumed Group} \\
	& & S & V & C & B & K & E & M & Met & P & S & C & B & E \vspace{0.3cm} \\ \hline \\[-1,5em] \hline \vspace{-0.2cm} \\
	$\Delta d_{g\phi}$ & mean & -0.24 & -0.29 & 0.63 & 0.10 & 0.46 & 1.15 & 1.18 & 2.14 & -0.94 & -0.24 & 1.06 & -1.00 & 1.40 \\
	                   & s.d. & 1.37 &  2.17 & 1.12 & 1.10 & 1.03 & 1.42 & 1.20 & 0.92 & 1.65 & 1.40 & 1.16 & 2.01 & 1.40 \\
	                  & count & 146 & 17 & 49 & 32 & 26 & 8 & 8 & 7 & 23 & 207 & 71 & 42 & 15\vspace{0.2cm} \\
\hline
\end{tabular}
\end{center}
\end{table}

\afterpage{
\begin{figure}
  \centering
  \begin{subfigure}[b]{.47\linewidth}
    \includegraphics[width=\linewidth]{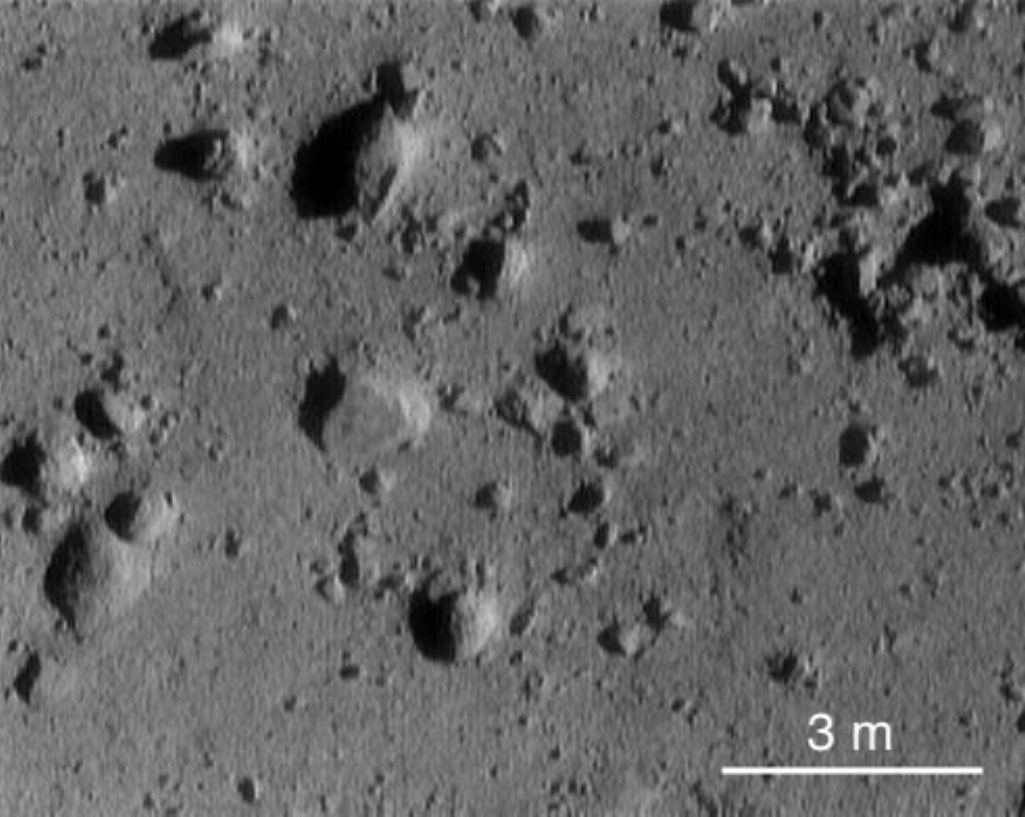}
    \caption{Eros}
  \end{subfigure}
  \begin{subfigure}[b]{.479\linewidth}
    \includegraphics[width=\linewidth]{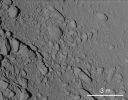}
    \caption{Itokawa}
  \end{subfigure} \\
  \begin{subfigure}[b]{.474\linewidth}
    \includegraphics[width=\linewidth]{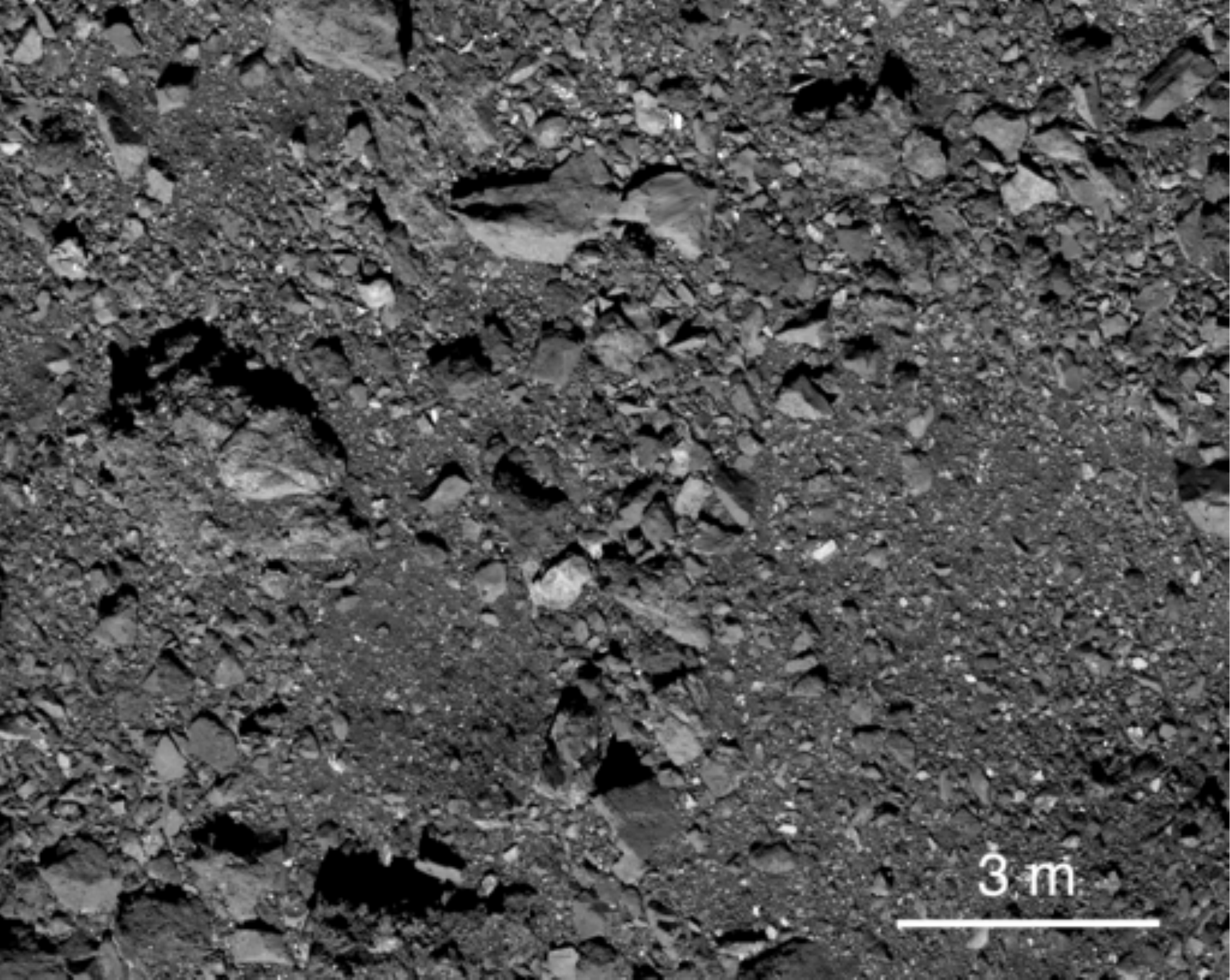}
    \caption{Bennu}
  \end{subfigure}
  \begin{subfigure}[b]{.474\linewidth}
    \includegraphics[width=\linewidth]{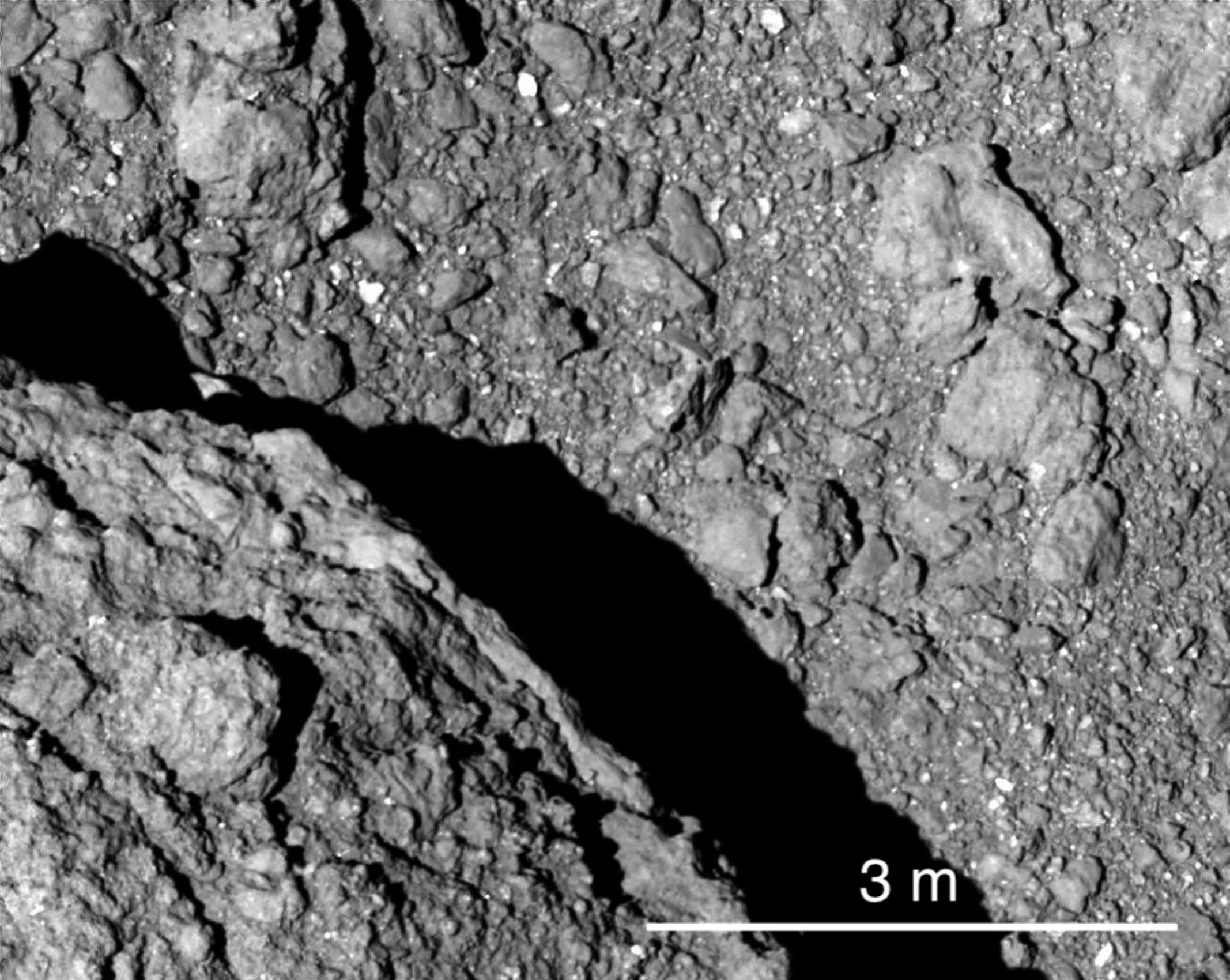}
    \caption{Ryugu}
  \end{subfigure}
  \caption{Spacecraft images of the regolith on Eros, Itokawa, Bennu, and Ryugu acquired by the NEAR, Hayabusa, OSIRIS-REx, and Hayabusa 2 missions respectively.}
  \label{fig:ErosIto}
\end{figure}

\begin{figure}[p!]
  \centering
  \renewcommand{\arraystretch}{0.2}
  \begin{tabular}[h!]{c}
	\includegraphics[clip,trim = 0.0cm 1.3cm 0cm 0.2cm,width=.85\linewidth]{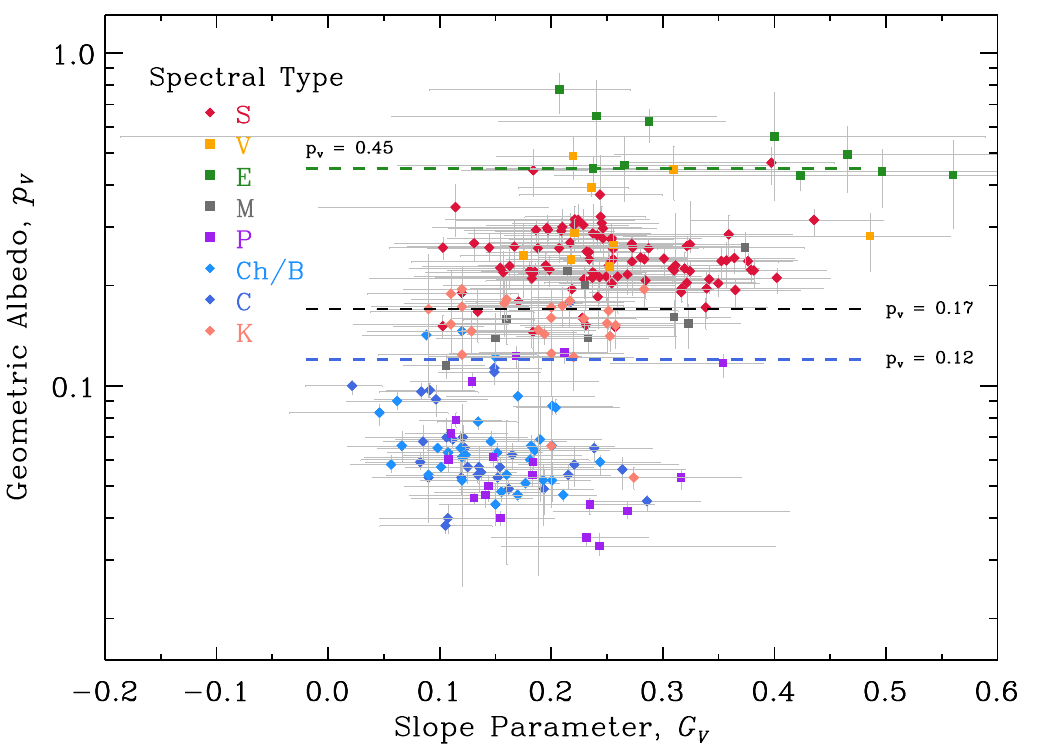} \\
	\includegraphics[clip,trim = 0.0cm 0.1cm 0cm 0.2cm,width=.85\linewidth]{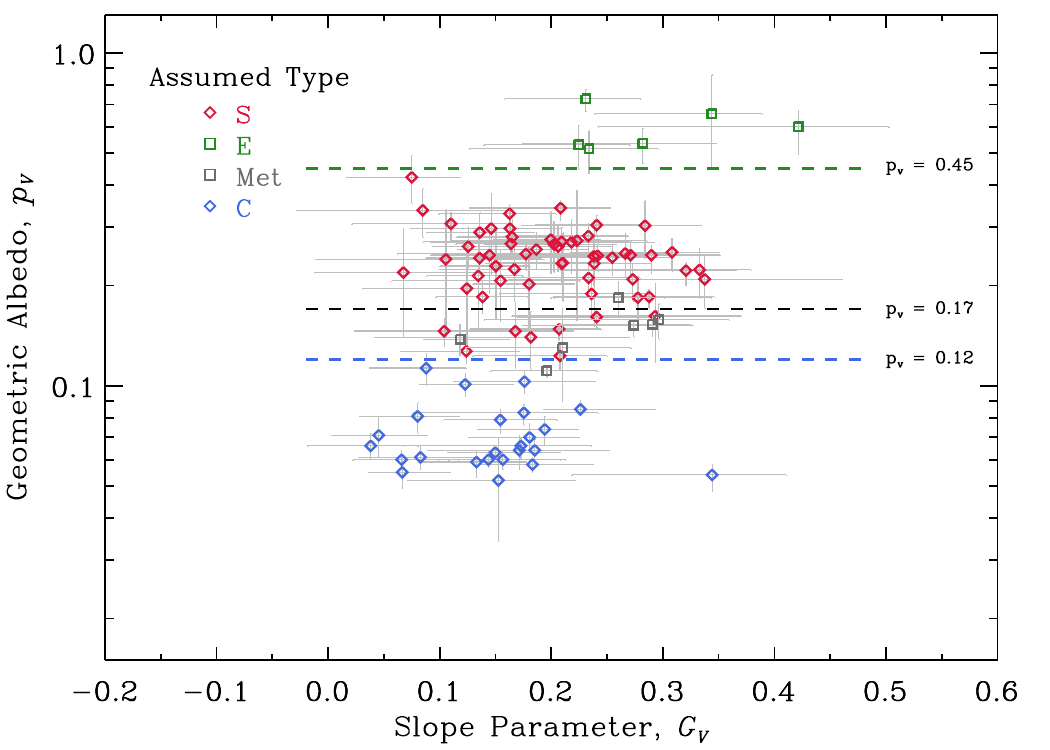}
  \end{tabular}
  \caption{Geometric albedo ($p_V$) versus slope parameter ($G_V$) grouped by spectral type for all asteroids in this study. The top panel (filled symbols) shows the objects classified on the basis of spectral or color data and family membership, when applicable. The bottom panel (open symbols) are the objects classified only on the basis of albedo.}\label{fig:albG}
\end{figure}

\begin{figure}[pb!]
\centering
	\begin{tabular}{c}
		\includegraphics[clip,trim = 0.1cm 0.05cm 0.3cm 0.2cm,width=.8\linewidth]{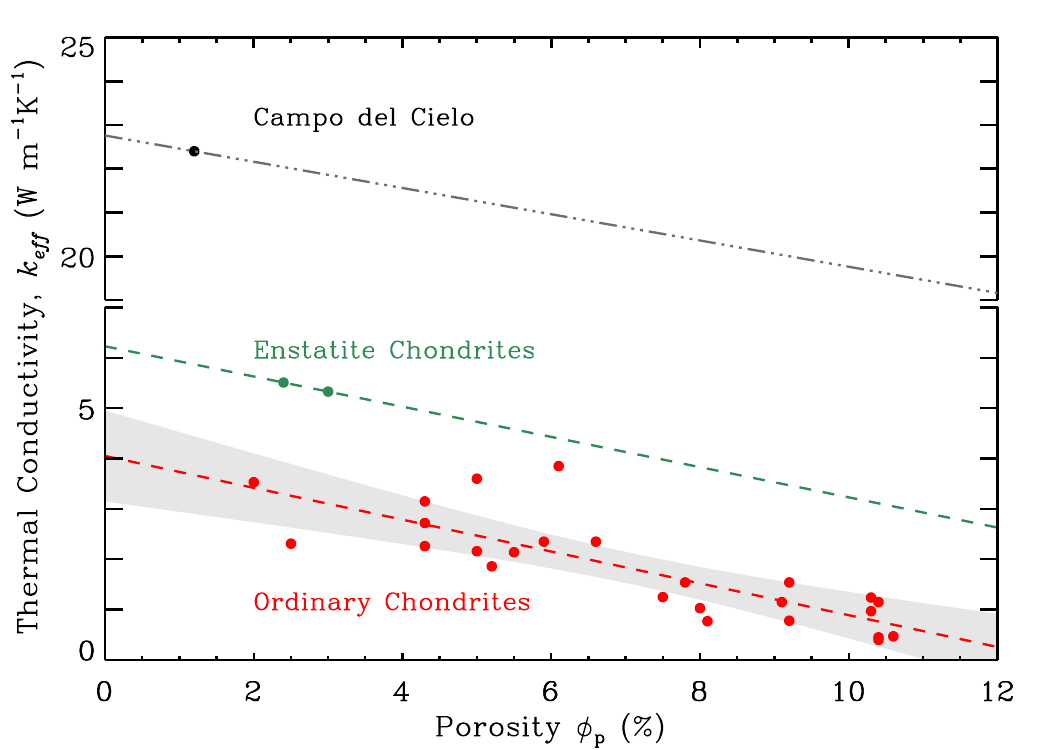} \\
		\includegraphics[clip,trim = 0.1cm 0.2cm 0.3cm 0.2cm,width=.8\linewidth]{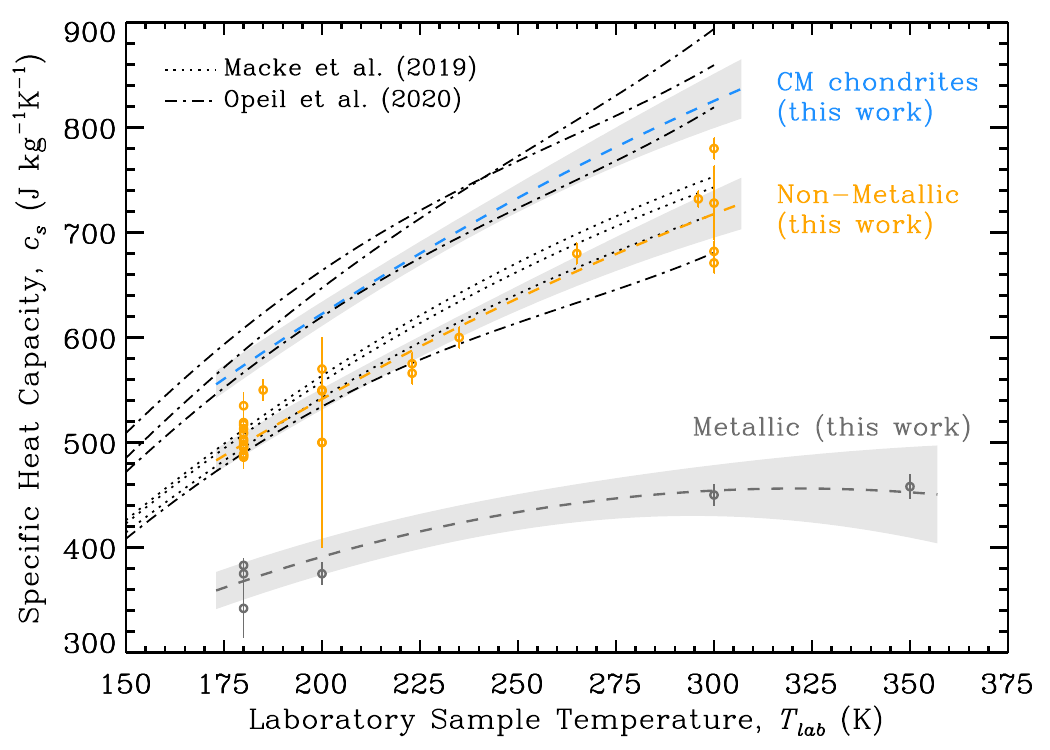}
	\end{tabular}
	\caption{Meteorite thermal conductivity as a function of porosity (top; \autoref{tab:porosityk}) and specific heat capacity as a function of temperature (bottom; \autoref{tab:heatcapT}). Dashed lines from this work show linear fits in the top panel, and parabola fits through the origin in the bottom panel, with grey regions representing 95\% confidence to the fits. The dashed-dotted line through the Campo del Cielo datum in the top panel assumes a slope of 0.3. Note the break in the y-axis in the top panel.}\label{fig:kheatcap} 
\end{figure}


\begin{figure}[p!]
  \centering
  \begin{tabular}[b]{c}
	\includegraphics[clip,trim = 0cm 1.45cm 0cm 0cm,width=.6\linewidth]{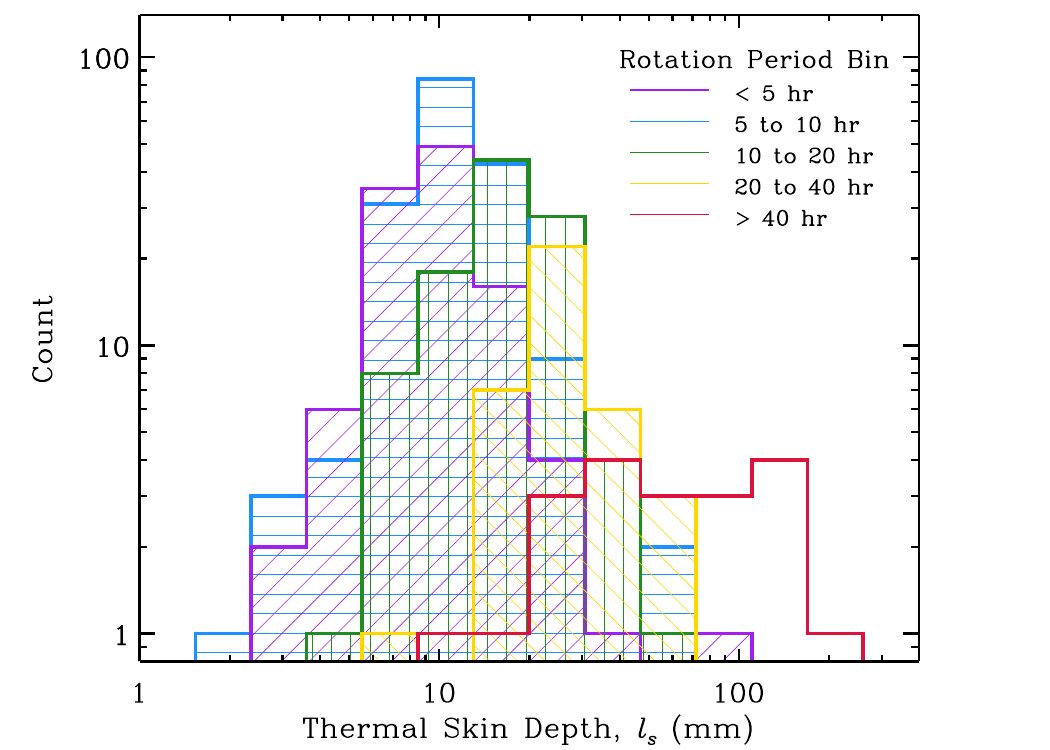} \\
	\includegraphics[clip,trim = 0cm 1.45cm 0cm 0cm,width=.6\linewidth]{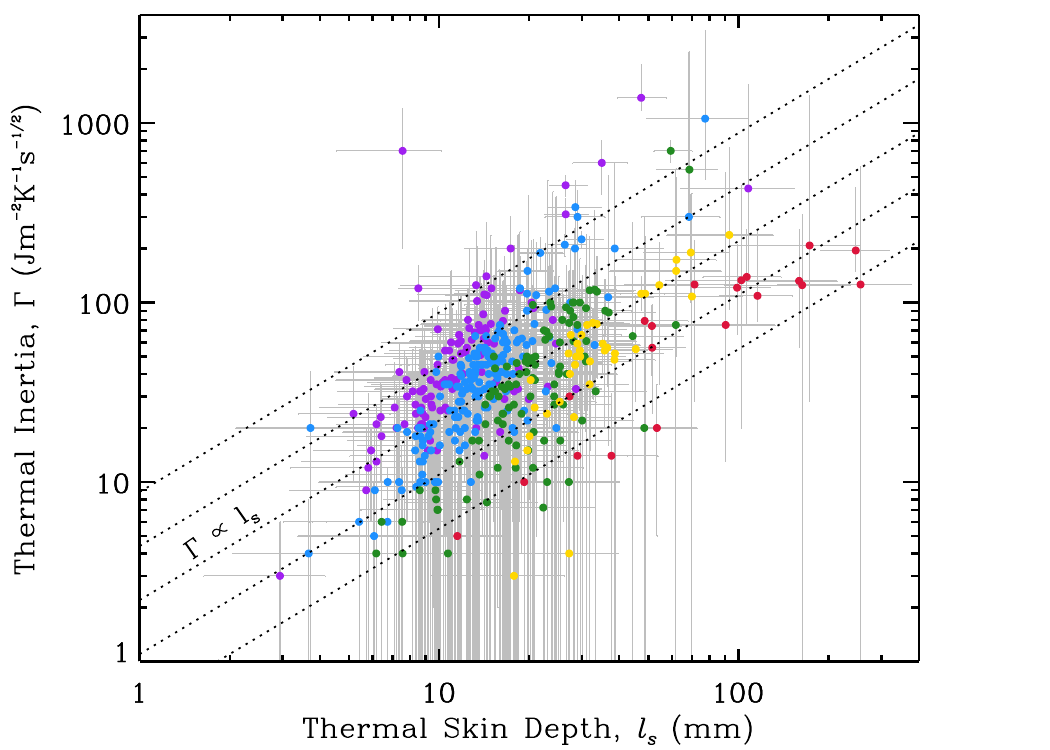} \\
	\includegraphics[clip,width=.6\linewidth]{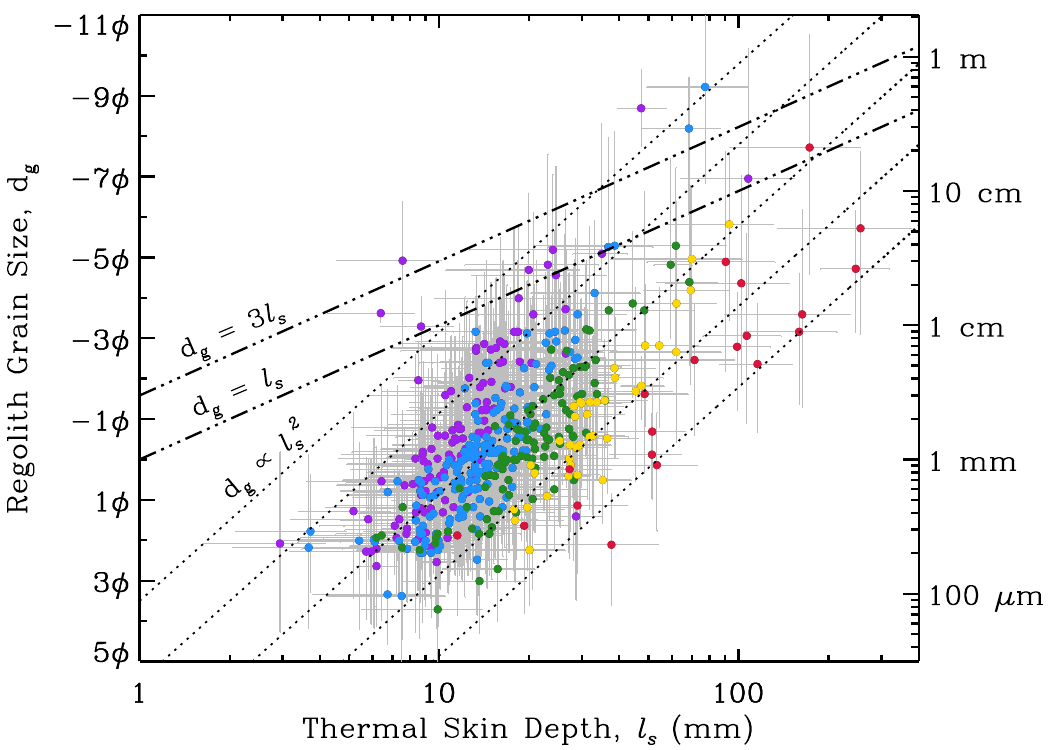}
  \end{tabular}
  \caption{{\it Top:} Histogram distributions of thermal skin depths, grouped by rotation period. Thermal inertia ({\it middle}) and regolith grain size {\it bottom} as a function of thermal skin depth, with colors indicating rotation period bin defined in the top panel. Dashed lines in the middle and bottom panels show proportionality relationships between the skin depth and thermal inertia or grain size, as shown in each figure. The dash-dot-dot lines indicate where the grain size is equal to the skin depth and three times the skin depth.} \label{fig:skindepth}
\end{figure}

\begin{landscape}

\begin{figure}[pb!]
  \centering
  \begin{tabular}[pb]{c}
	\includegraphics[clip,trim = 0.1cm 0.07cm 0.07cm 0.2cm,width=.5\linewidth]{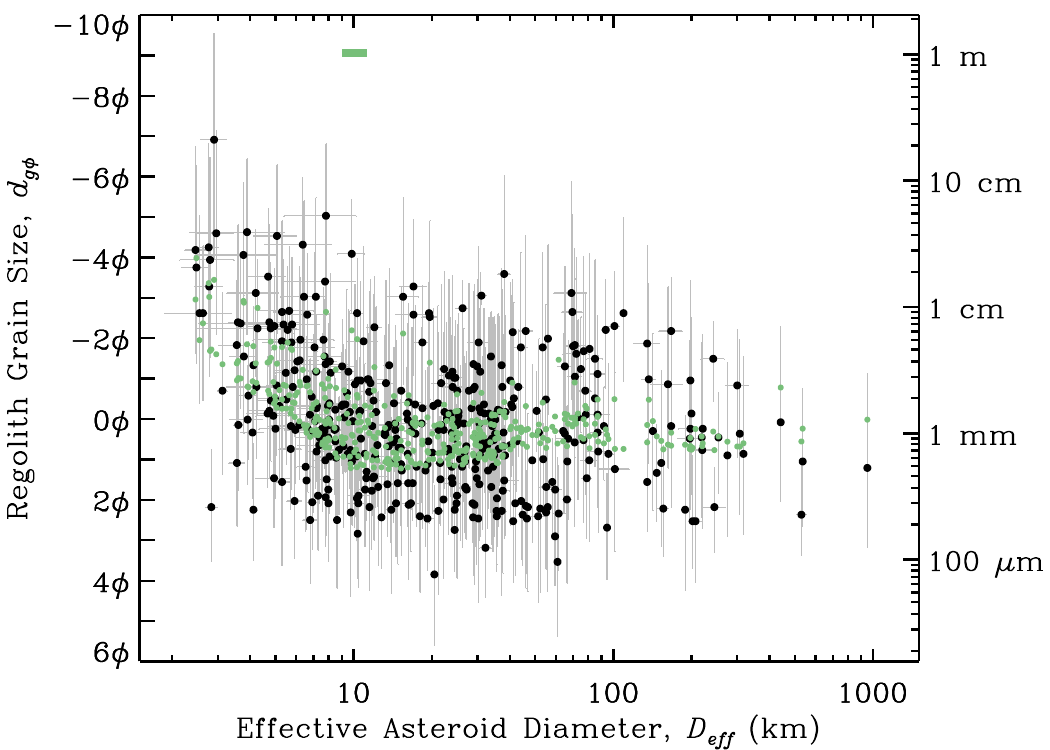}\includegraphics[clip,trim = 0.6cm 0.07cm 0.07cm 0.2cm,width=.485\linewidth]{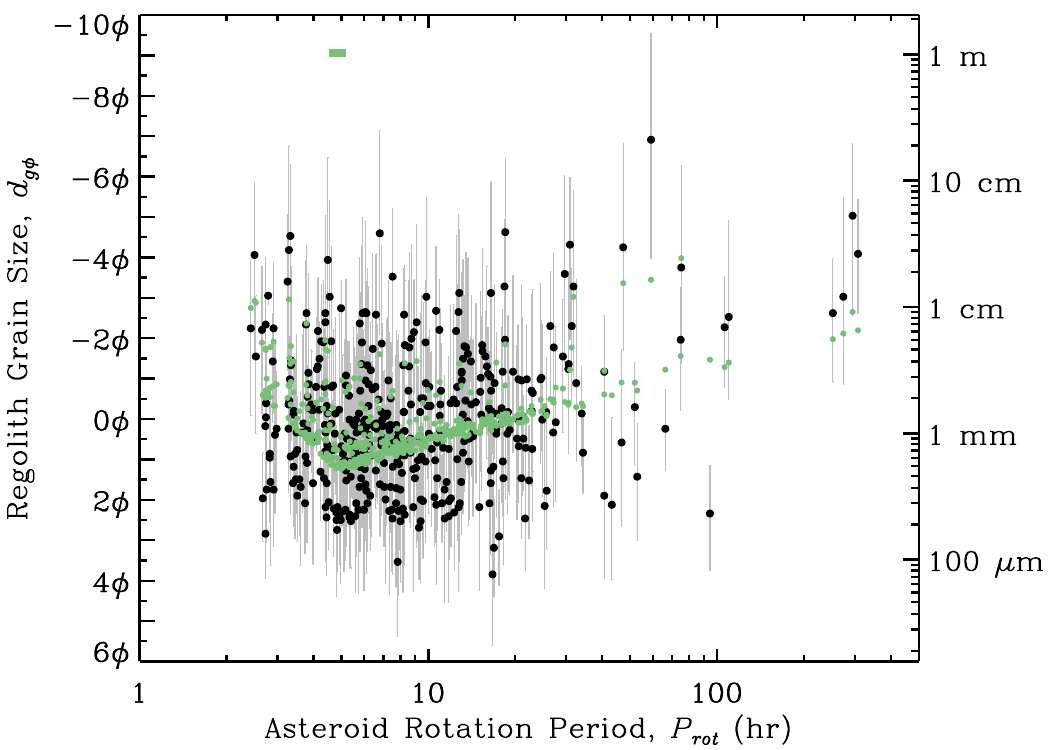}
  \end{tabular}
  \caption{Multi-linear model fit {\it M-4}, shown by colored points, and the grain size dataset, shown by black dots, as a function of asteroid diameter (left) and rotation period (right). The colored bars at the top of each panel indicate the 1$\sigma$ range in uncertainty in the break-point between the two segments.}\label{fig:modelfit}
\end{figure}
\end{landscape}

\afterpage{
\begin{figure}[pb!]
    \includegraphics{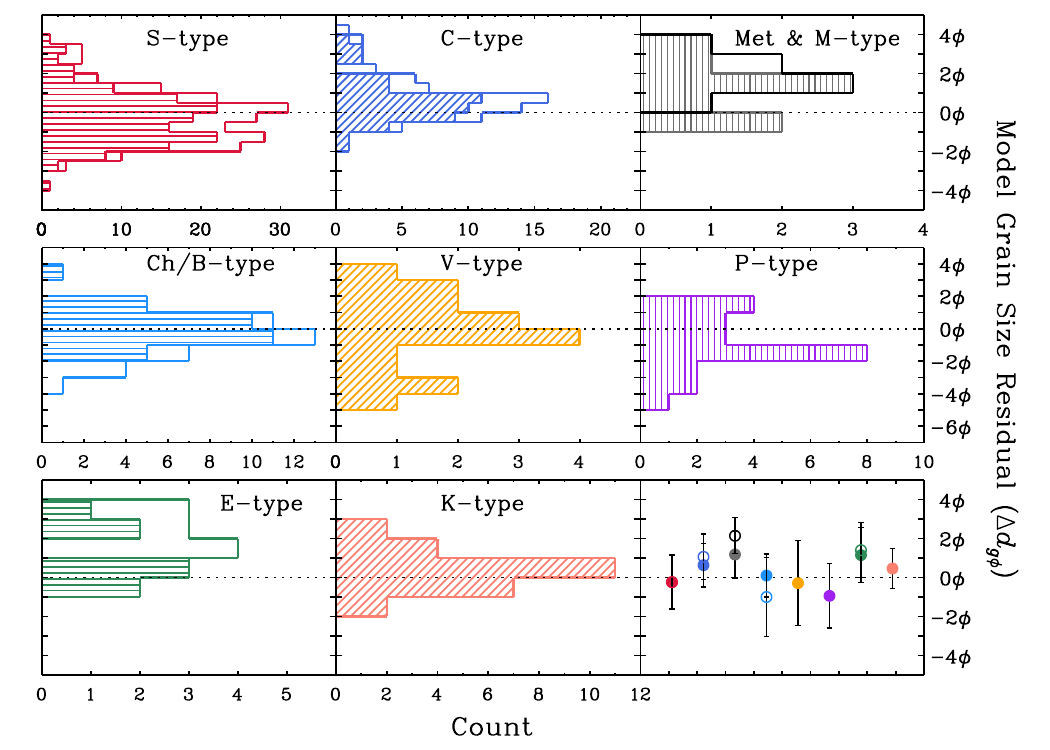}
  \caption{Grain size model residual distributions for different compositional groups and the group means and standard deviations (bottom right panel). Unfilled histograms and open circles in the bottom right panel indicate that objects with assumed spectral classification are included. The filled grey histogram and the black line in the upper right panel are for M-types and metal-rich asteroids, respectively. The dotted horizontal line at 0$\phi$ in each panel is shown for comparison purposes. Note the change in the y-axis range in the middle three panels.}\label{fig:residcomp}
\end{figure}
}


\afterpage{
\begin{landscape}
\pagestyle{mylandscape}
\begin{figure}[p!]
  \centering
  \setlength{\tabcolsep}{-1.3pt}
  \begin{tabular}[h!]{lr}
	\includegraphics[clip,trim = 0.1cm 0.05cm 2cm 0.2cm,height=.33\linewidth]{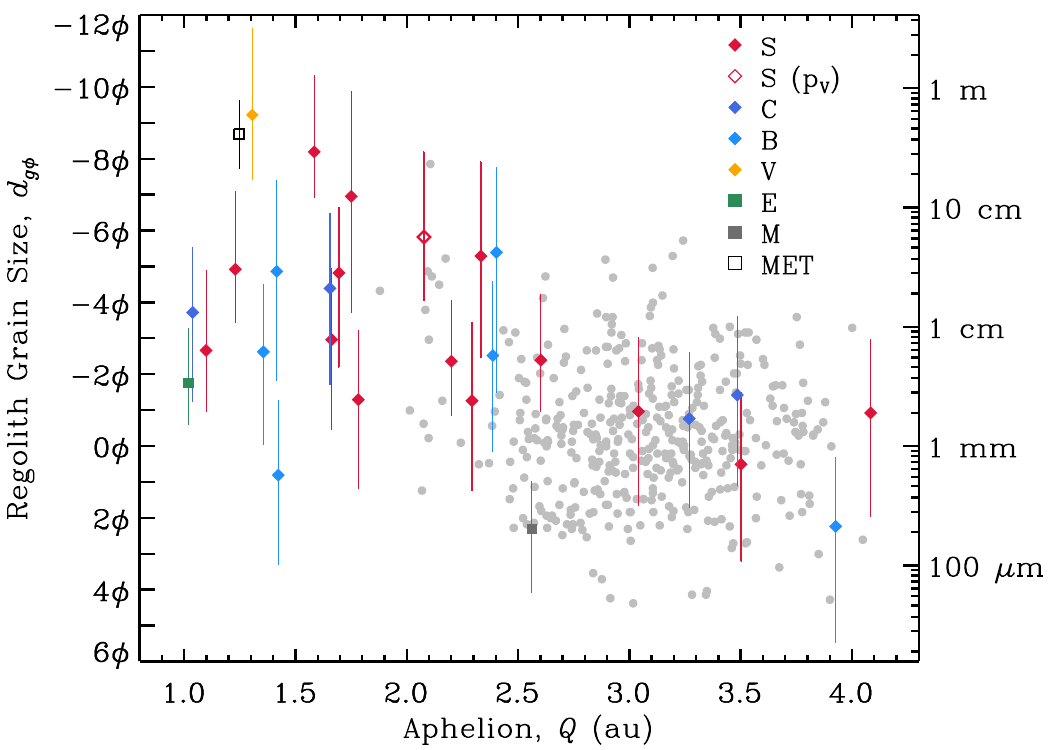} &
	\includegraphics[clip,trim = 2.2cm 0.05cm 0.12cm 0.2cm,height=.33\linewidth]{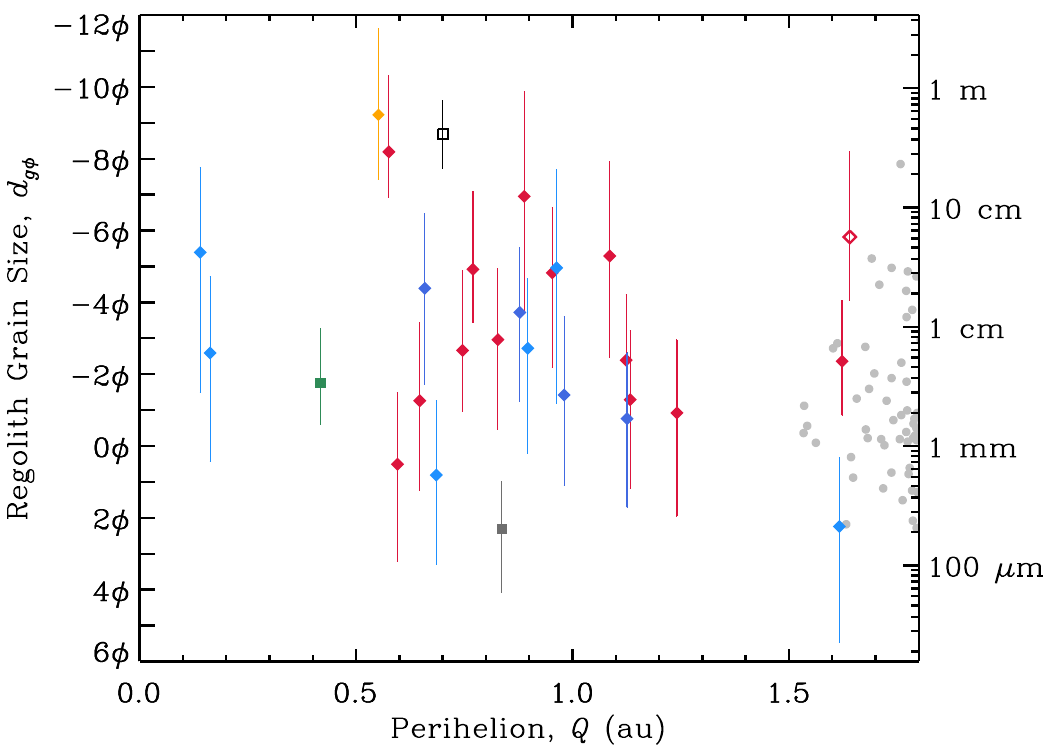} \\
	\includegraphics[clip,trim = 0.1cm 0.05cm 2cm 0.2cm,height=.33\linewidth]{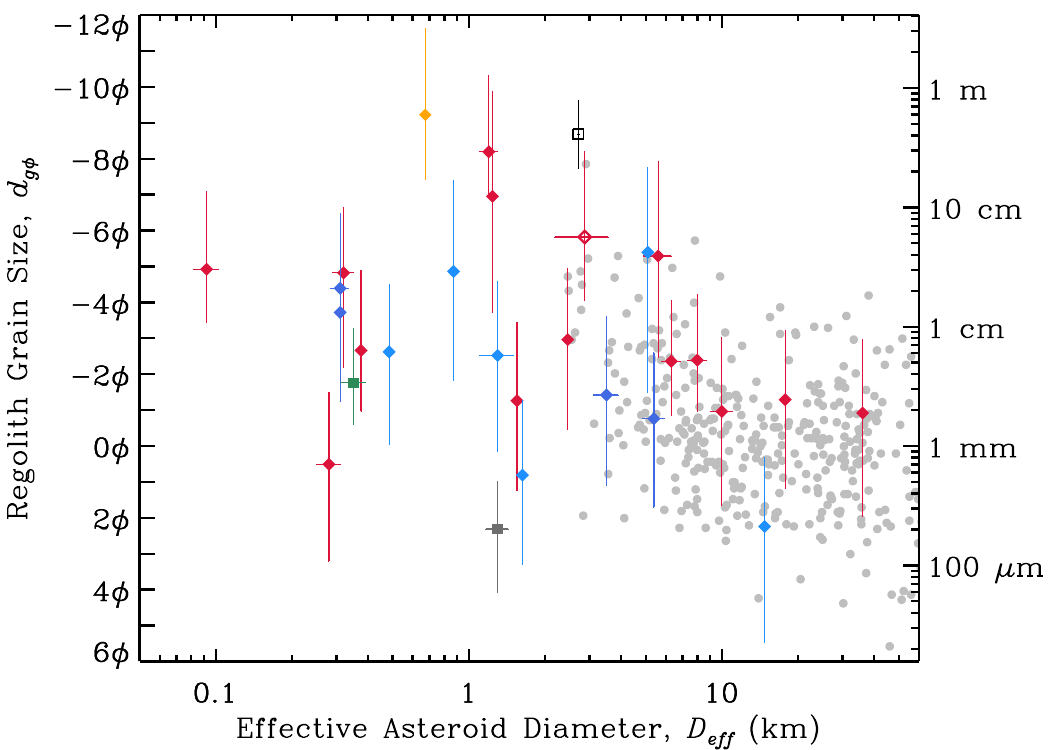} &
	\includegraphics[clip,trim = 2.3cm 0.05cm 0.12cm 0.2cm,height=.33\linewidth]{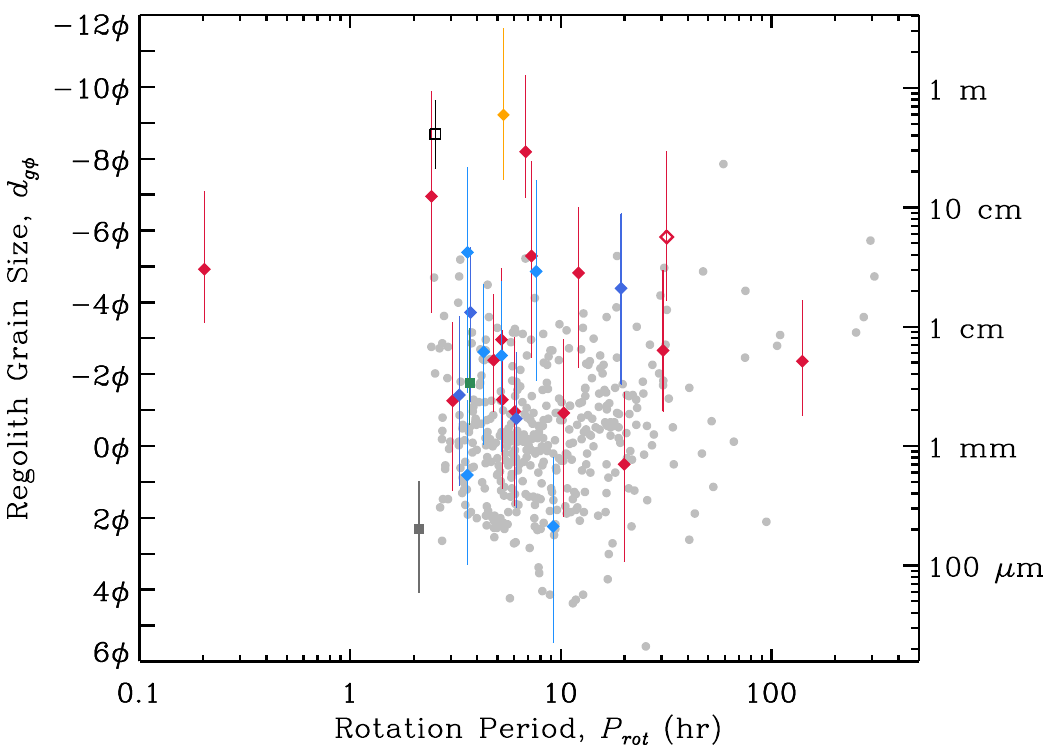}
  \end{tabular}
  \caption{Regolith grain sizes of NEAs as a function of a) aphelion distance, b) perihelion distance, c) asteroid diameter, and d) rotation period. Symbol colors indicate spectral type as noted in the upper left panel, and are consistent with previous figures. The open symbols are objects with an inferred spectral type (see text) and grey dots are MBAs.}\label{fig:NEAgrainsize}
\end{figure}
\end{landscape}
}
}

\end{document}

%% file: grainresults_finfix2.txt
(1) Ceres &   951 $\pm$ 8 &   0.100$^{+0.004}_{-0.006}$ &   25$^{+15}_{-10}$ &   \hyperlink{TI1}{[T1]} &   C\hyperlink{T2}{$^2$} &   9.0742  &   238 $\pm$ 15 &   C$^{S}$ &   0.91$^{+2.10}_{-2.47}$ &   1.31$^{+0.67}_{-0.77}$ & 12 $\pm$ 3.3 \\
(2) Pallas &   536 $\pm$ 5 &   0.142$^{+0.006}_{-0.005}$ &   30$^{+15}_{-15}$ &   \hyperlink{TI1}{[T1]} &   B\hyperlink{T2}{$^2$} &   7.8132\hyperlink{TTdag}{$^\dagger$}  &   234 $\pm$ 15 &   B$^{S}$ &   0.74$^{+1.70}_{-1.92}$ &   1.24$^{+0.60}_{-0.73}$ & 13.2 $\pm$ 3.7 \\
(3) Juno &   254 $\pm$ 4 &   0.209$^{+0.020}_{-0.019}$ &   70$^{+30}_{-40}$ &   \hyperlink{TI1}{[T1]} &   Sk\hyperlink{T2}{$^2$} &   7.2095  &   245 $\pm$ 15 &   S$^{S}$ &   0.11$^{+1.47}_{-1.17}$ &   -0.19$^{+1.13}_{-1.43}$ & 16 $\pm$ 4.7 \\
(4) Vesta &   530 $\pm$ 24 &   0.394$^{+0.011}_{-0.024}$ &   30$^{+10}_{-10}$ &   \hyperlink{TI2}{[T2]} &   V\hyperlink{T2}{$^2$} &   5.3421  &   230 $\pm$ 10 &   V$^{S}$ &   2.14$^{+1.07}_{-1.40}$ &   1.84$^{+0.40}_{-0.63}$ & 10 $\pm$ 2.3 \\
(6) Hebe &   198 $\pm$ 3 &   0.240$^{+0.010}_{-0.010}$ &   50$^{+40}_{-35}$ &   \hyperlink{TI3}{[T3]} &   S\hyperlink{T2}{$^2$} &   7.2744\hyperlink{TTdag}{$^\dagger$}  &   205 $\pm$ 15 &   S$^{S}$ &   -1.39$^{+2.10}_{-2.63}$ &   -1.49$^{+1.00}_{-1.20}$ & 16.3 $\pm$ 5.8 \\
(8) Flora &   142 $\pm$ 2 &   0.252$^{+0.015}_{-0.014}$ &   50$^{+35}_{-30}$ &   \hyperlink{TI1}{[T1]} &   S\hyperlink{T2}{$^2$} &   12.865  &   244 $\pm$ 15 &   S$^{S}$ &   -0.06$^{+2.10}_{-2.57}$ &   -0.22$^{+1.20}_{-1.43}$ & 19.7 $\pm$ 6.4 & Flora \\